\documentclass[ALICE,manyauthors]{cernphprep}

\usepackage[comma,square,numbers,sort&compress]{natbib}
\usepackage{hyperref}
\usepackage{lineno}
\usepackage{multirow}
\usepackage{color}
\usepackage{rotating}
\usepackage{graphics}
\usepackage{soul}
\usepackage{bm}
\usepackage{hyperref}
\usepackage{float}

\usepackage[T1]{fontenc}

\newcommand{\sqrtS}{\ensuremath{\sqrt{s}}}

\newcommand{\sqrtSnn}{\ensuremath{\sqrt{s_{\mathrm{NN}}}}}

\newcommand{\pT}{\ensuremath{p_{\mathrm{T}}}}

\newcommand{\BpT}{\ensuremath{\bf{p}_{\mathrm{\bf{T}}}}}
\newcommand{\meanpT}{$\langle$\ensuremath{p_{\mathrm{T}}}$\rangle$}
\newcommand {\dedx}{d$E$/d$x$}

\newcommand{\dndy}{d$N$/d$y$}
\newcommand{\Bdndy}{\bf{d}$\bf{N}$/\bf{d}$\bf{y}$}
\newcommand{\BmeanpT}{$\langle$\ensuremath{\bf{p}_{\mathrm{\bf{T}}}}$\rangle$}

\newcommand{\modrap} {$\left | y \right |$}
\newcommand{\modeta} {$\left | \eta \right |$}

\newcommand{\pp}{\ensuremath{\mathrm {p\kern-0.05em p}}}
\newcommand{\PbPb}{\ensuremath{\mbox{Pb--Pb}}}
\newcommand{\pPb}{\ensuremath{\mbox{p--Pb}}}

\newcommand{\pip}{\ensuremath{\pi^{+}}}
\newcommand{\pim}{\ensuremath{\pi^{-}}}
\newcommand{\ppm}{\ensuremath{\pi^{\pm}}}
\newcommand{\pmp}{\ensuremath{\pi^{\mp}}}

\newcommand{\VZERO}{V$^{\mathrm 0}$}

\newcommand {\gmom} {GeV/$c$}

\newcommand {\gmass} {GeV/$c^{\mathrm 2}$}
\newcommand {\fmc} {fm/$c$}
\newcommand {\Luminosity}{$\mathcal{L}_{\mathrm{int}}$}
\newcommand{\K}{\mbox{$\mathrm {K}$}}
\newcommand{\Kp}{\mbox{$\mathrm {K}^+$}}
\newcommand{\Km}{\mbox{$\mathrm {K}^-$}}
\newcommand{\Kpm}{\mbox{$\mathrm {K}^{\pm}$}}

\newcommand{\pion}{\mbox {$\mathrm {\pi}$}}
\newcommand{\pionp}{\mbox {$\mathrm {\pi}^+$}}
\newcommand{\pionm}{\mbox {$\mathrm {\pi}^-$}}

\newcommand{\kshort}{\mbox{$\mathrm {K^0_S}$}}
\newcommand{\Bkshort}{\mbox{$\mathrm {\bf{K^0_S}}$}}
\newcommand{\kzero}{\mbox{$\mathrm {K^0}$}}
\newcommand{\rmLambda}{\mbox{$\mathrm {\Lambda}$}}

\newcommand{\rmXi}{\mbox{$\mathrm {\Xi^{-}}$}}

\newcommand{\kstarZ}{\mbox{\K$^{*}$(892)$^{\mathrm{0}}$}}
\newcommand{\akstarZ}{\mbox{$\overline{\K^{*}(892)^\mathrm{0}}$}}
\newcommand{\kstarch}{\mbox{\K$^{*}$(892)$^{\mathrm{\pm}}$}}
\newcommand{\kstarchp}{\mbox{\K$^{*}$(892)$^{\mathrm{+}}$}}
\newcommand{\kstarchn}{\mbox{\K$^{*}$(892)$^{\mathrm{-}}$}}
\newcommand{\simplekstarch}{\mbox{\K$^{*\mathrm{\pm}}$}}
\newcommand{\simplekstarp}{\mbox{\K$^{*\mathrm{+}}$}}
\newcommand{\simplekstarn}{\mbox{\K$^{*\mathrm{-}}$}}
\newcommand{\Bsimplekstarch}{\mbox\bf{K$^{\mathbf{*{\pm}}}$}}
\newcommand{\simplekstarZ}{\mbox{\K$^{*\mathrm{0}}$}}
\newcommand{\kstar}{\mbox{\K$^{*}$}}
\newcommand{\kstarPDG}{\mbox{\K$^{*}$(892)}}

\newcommand{\rsimplekstar}{\mbox{\K$^{*}$}}

\newcommand{\sbar} {{$\overline{\mathrm s}$}}
\newcommand{\ubar} {{$\overline{\mathrm u}$}}
\newcommand{\dbar} {{$\overline{\mathrm d}$}}

\begin{document}

%

%
\begin{titlepage}
\PHyear{2021}
\PHnumber{086}   
\PHdate{10 May}  

\title{Measurement of \kstarch~production in inelastic pp collisions at the LHC}
\ShortTitle{Measurement of \kstarch~production}   

\Collaboration{ALICE Collaboration\thanks{See Appendix~\ref{app:collab} for the list of collaboration members}}
\ShortAuthor{ALICE Collaboration} 

\begin{abstract}
The first results on \kstarch~resonance production in inelastic pp collisions at LHC energies of \sqrtS~=~5.02, 8, and 13~TeV are presented. The \kstarch~has been reconstructed via its hadronic decay channel 
\kstarch $\rightarrow$ \kshort $~+~\pion^{\pm}$ with the ALICE detector. Measurements of transverse momentum distributions, \pT-integrated yields, and mean transverse momenta for charged \kstarPDG~are found to be consistent with previous ALICE measurements for neutral \kstarPDG~within uncertainties. 
For \pT~$>$~1~\gmom~the \kstarch~transverse momentum spectra become harder with increasing centre-of-mass energy from 5.02 to 13~TeV, similar to what previously observed for charged kaons and pions. For \pT~$<$~1~\gmom~the \kstarch~yield does not evolve significantly and the abundance of \kstarch~relative to \K~is rather independent of the collision energy. The transverse momentum spectra, measured for \kstarch~at midrapidity in the interval 0~$<$~\pT~$<$~15~\gmom, are not well described by predictions of different versions of PYTHIA 6, PYTHIA 8 and EPOS-LHC event generators. These generators reproduce the measured \pT-integrated \simplekstarch/\K~ratios and describe well the momentum dependence for 
\pT~$<$~2~\gmom.
\end{abstract}
\end{titlepage}

\setcounter{page}{2} 


\section{Introduction}
\label{sec:intro}

Measurements of identified hadron production in high-energy proton-proton interactions provide key observables to characterize the global properties of the collisions. Particle production at high collider energies originates from the interplay of perturbative (hard) and non-perturbative (soft) Quantum Chromodynamic (QCD) processes. 
Soft scattering processes and parton shower hadronization dominate the bulk of particle production at low transverse momenta and can only be modeled phenomenologically. 

At the Large Hadron Collider (LHC)~\cite{LHC}, the small Bjorken $x$ regime is probed and contributions from hard-scattering processes  are more relevant with increasing centre-of-mass energy. This produces a hardening of the transverse momentum spectra, as already observed in Refs.~\cite{pp13TeV,ALICE_particle_13TeV}. 
Measurements of strange hadrons such as the \kstarPDG~vector meson at different collision energies allow for testing and tuning perturbative QCD and low-transverse momentum phenomenological calculations~\cite{PYTHIA6,PYTHIA8,EPOS}, including strangeness production.

In the following, \simplekstarZ~denotes \kstarZ~and \akstarZ, \simplekstarch~stands for \kstarchp~and \kstarchn, while \kstar~indicates \simplekstarZ~and \simplekstarch.

In heavy-ion collisions, due to their short lifetimes  
comparable with the lifetime of the hadronic phase of the system~\cite{HBT}, resonances such as \kstar~($\tau \approx$~4~\fmc) are sensitive probes of the dynamical evolution of the fireball.
Re-scattering and regeneration in the hadron gas may change the number of resonances reconstructed via the hadronic decay channels compared to those predicted by thermal models at the chemical freeze-out, i.e. when the inelastic interactions stop. 
The \rsimplekstar~vector meson and its corresponding ground state, the \K, have an identical quark content. They differ only in mass, lifetime and relative orientation of their quark spins. Therefore, the
\kstar$/$\K~ratio is an ideal observable to study the \kstar~properties and the freeze-out conditions in relativistic heavy-ion collisions. The integrated yield ratio \simplekstarZ/\K~exhibits a suppression with respect to pp collisions, which increases with the centrality of the collisions~\cite{STAR_Kstar_200,Kstar_PbPb,Kstar_highpt,ALICE_Kstar_5TeV}.
 This could be explained as due to the dominance of re-scattering effects of \simplekstarZ~decay products over regeneration processes in the hadronic phase of the collisions.

Hints of the suppression of \simplekstarZ/\K~were observed also in high-multiplicity \pPb~and pp collisions~\cite{Kstar_pPb,ALICE_7TeV_multiplicity,Kstar_phi_multiplicity_13TeV} at LHC energies, suggesting the possible presence of re-scattering effects and thus of a hadronic phase with a short but finite lifetime in small collision systems. 
The observed multiplicity-dependent suppression should therefore be validated by measurements with an increased precision. This is particularly important for small systems such as pp and \pPb~because the \simplekstarZ/\K~ratios, measured in the highest and lowest multiplicity event classes differ by less than 2$\sigma$~\cite{Kstar_pPb,ALICE_7TeV_multiplicity,Kstar_phi_multiplicity_13TeV}, with the largest uncertainty in the ratio being relative to the \simplekstarZ~yield measurement.
In this work, the \kstar/\K~ratio is studied with increased precision by measuring the production yield of \simplekstarch~in pp collisions with the ALICE detector~\cite{ALICE_JINST}. 
The production of charged and neutral \kstar~vector mesons is expected to be comparable. Indeed, they have a similar quark composition, \kstarchp~=~(u\sbar), \kstarZ~=~(d\sbar), 
\kstarchn~=~(\ubar s) and \akstarZ~=~(\dbar s), and their masses differ by about 0.004~\gmass, being 
M(\simplekstarch)~=~0.89166~$\pm$~0.0026~\gmass~\cite{PDG} and M(\simplekstarZ)~=~0.89581~$\pm$~0.0019~\gmass~\cite{PDG}.  
At LHC energies, the measurement of the \simplekstarch~and \simplekstarZ~strange vector mesons is quite challenging. These are reconstructed via their hadronic decay into a charged pion and a kaon: a neutral kaon for \simplekstarch~and a charged kaon for \simplekstarZ. Because of the different strategies used for their identification in ALICE, \kshort~are measured with a lower systematic uncertainty than charged kaons~\cite{ALICE_7TeV_multiplicity,ALICE_particle_13TeV}.

In this paper, transverse momentum (\pT) distributions of \simplekstarch~resonances at midrapidity (\modrap~$<$~0.5) are presented for the first time for inelastic pp collisions at the LHC. 
The evolution of the \pT~distributions with the energy was investigated by studying pp collisions at the centre-of-mass energies of \sqrtS~$=$~5.02, 8, and 13~TeV. 
The similarity of the charged and neutral \kstar~production was checked by comparing \simplekstarch~results with existing  \simplekstarZ~measurements at the same collision energy~\cite{ALICE_Kstar_5TeV,ALICE_Kstar_8TeV,ALICE_particle_13TeV}.  
These measurements are a useful probe of strangeness production and provide input to tune Monte Carlo event generators such as PYTHIA and EPOS-LHC~\cite{PYTHIA6,PYTHIA8,EPOS} as a function of collision energy. Furthermore, the measurements in inelastic pp collisions at \sqrtS~$=$~5.02, 8, and 13~TeV reported in this paper serve as reference data to study nuclear effects in \pPb~and \PbPb~collisions.
  
The paper is organized as follows. In Sec.~\ref{sec:setup} the ALICE experimental setup is described, focusing on the detectors employed in the analysis presented here. Details on the event, track and particle identification as well as on the corrections applied to the measured raw yields and estimation of systematic uncertainties are discussed in Sec.~\ref{sec:data}. 
In Sec.~\ref{sec:results}, the results on the production of \simplekstarch~resonances are shown. These include the transverse momentum spectra, the mean transverse momenta, the per-event~\pT-integrated particle yields and the \simplekstarch/\K~=~(\simplekstarp~+~\simplekstarn)/(\K$^+$~+~\K$^-$) ratio as a function of the collision energy. All these observables are compared with similar results for \simplekstarZ. The comparison of the \pT~spectra with different event generator (PYTHIA6, PYTHIA8 and EPOS-LHC) predictions is also presented. 
In Sec.~\ref{sec:conclusions} results are summarized and conclusions are drawn.

\section{Experimental setup}
\label{sec:setup}

A detailed description of the ALICE detector and its performance can be found in Refs.~\cite{ALICE_JINST, ALICE_Performance}. The sub-detectors used for the analysis presented in this paper are the Inner Tracking System (ITS)~\cite{ALICE_JINST}, the Time Projection Chamber (TPC)~\cite{TPC}, and the V0 detectors~\cite{VZERO}. All tracking detectors are positioned in a solenoidal magnetic field B~=~0.5~T parallel to the LHC beam axis.  

Charged particle tracks are reconstructed by the ITS and the TPC. The ITS is the innermost barrel detector consisting of six cylindrical layers of high-resolution silicon tracking detectors. The innermost layers consist of two arrays of hybrid Silicon Pixel Detectors (SPD) located at an average radial distance $r$ of 3.9 and 7.6~cm from the beam axis and covering \modeta~$<$~2.0 and \modeta~$<$~1.4, respectively. 
The SPD is used to reconstruct the primary vertex (PV) of the collisions, which is found as a space point to which the maximum number of tracklets (track segments defined by pairs of points, one point in each SPD layer) converges. The outer layers of the ITS are composed of two layers of silicon drift and two layers of silicon strip detectors, with the outermost layer positioned at $r$~=~43~cm. 
The TPC is the main tracking device of ALICE. It is a large volume (90~m$^3$) cylindrical drift chamber with radial and longitudinal dimension of about 85~$<~r~<$~250~cm and $-250$~$<~z~<~$250~cm, respectively, covering for full-length tracks a pseudorapidity range of \modeta~$<$~0.9 over the full azimuth. The end-caps of the TPC are equipped with multiwire proportional chambers segmented radially into pad rows. Together with the measurement of the drift time, the TPC provides three dimensional space point information, with up to 159 samples per track. The resolution on the position is 1100-800~$\mu$m on the transverse plane and 1250-1100~$\mu$m along $z$.
 Charged tracks originating from the primary vertex can be reconstructed down to \pT~$\approx$~0.1~\gmom~\cite{ALICE_Performance}. The TPC enables charged particle identification (PID) via the measurement of the specific ionization energy loss (\dedx) with a resolution of about 5.2\%~\cite{ALICE_Performance} at low transverse momentum. 
A separation between \pion-\K~and \K-p at the level of two standard deviations is possible for 
\pT~$<$~0.8~\gmom~and 1.6~\gmom, respectively.

The V0 detectors are two forward scintillator hodoscopes employed for triggering and beam background suppression. They are placed along the beam axis on each side of the nominal interaction point (IP) at $z$~=~340~cm and $z=~-90$~cm, covering the pseudorapidity regions 2.8~$<~\eta~<$~5.1 (V0A) and $-3.7~<~\eta~<~-1.7$ (V0C), respectively.

The pp data at \sqrtS~=~5.02 and 13~TeV used in this paper were collected in 2015 while data at \sqrtS~=~8~TeV were collected in 2012. The data were collected with a minimum bias trigger requiring a hit in both V0 detectors, in coincidence with the arrival of proton bunches from both beam directions. 

The analysed data are low pile-up samples in which the average number of interactions per bunch crossing are $\mu~=$~0.019 $\pm$ 0.009, 0.02 $\pm$ 0.01 and 0.068 $\pm$ 0.003 for collisions at \sqrtS~=~5.02, 8, and 13~TeV, respectively.
Contamination from beam-gas events is removed offline by using timing information from the V0 detector, which has a time resolution better than 1~ns. The events in which pile-up or beam-gas interaction occurred are also rejected by exploiting the correlation between the number of SPD hits and the number of SPD tracklets, as discussed in detail in Ref.~\cite{ALICE_Performance}.

The events selected from the analysis are required to have a reconstructed primary vertex with its position along the beam axis being within 10~cm with respect to the nominal interaction point (the centre of the ALICE barrel). The events containing more than one reconstructed vertex are tagged as pile-up occurring within the same bunch crossing and discarded for the analysis.
 
The size of the analyzed samples after selection and the corresponding pp integrated luminosities are given in Tab.~\ref{Statistic}. In the same table, the primary vertex reconstruction efficiency $\epsilon_{\mathrm{vertex}}$~and the trigger selection efficiency 
$\epsilon_{\mathrm{trig}}$~are also reported. For each energy, the $\epsilon_{\mathrm{trig}}$ value, mainly defined by the charged particle multiplicity of the collision, is the ratio between the V0-triggered cross section~\cite{Luminosity_5TeV,Luminosity_8TeV,Luminosity_13TeV} and the inelastic cross section~\cite{Inelastic_cross_section} and the 
$\epsilon_{\mathrm{vertex}}$~is the fraction of V0-triggered events for which a primary vertex is reconstructed. 

\begin{table*} 
\caption{Number of minimum bias events after event selection (N$_{\mathrm {MB}}$), integrated luminosity (\Luminosity), the trigger selection efficiency ($\epsilon_{\mathrm{trig}}$), and the primary vertex reconstruction efficiency 
($\epsilon_{\mathrm{vertex}}$) for the analyzed data sets. The uncertainty on $\epsilon_{\mathrm{vertex}}$ is lower than 0.1$\%$.}
\begin{center}
\begin{tabular}{ccccc}
\hline\noalign{\smallskip}
\sqrtS~(TeV) & N$_{\mathrm {MB}}$~(10$^7$) & \Luminosity~(nb$^{-1}$) & $\epsilon_{\mathrm{trig}}$ & $\epsilon_{\mathrm{vertex}}$\\
\hline
\hline\noalign{\smallskip}
5.02 & 10.87 & 2.12 $\pm$ 0.05 & 0.757 $\pm$ 0.019 & 0.958\\
8.0 & 6.99 & 1.25 $\pm$ 0.03 & 0.772 $\pm$ 0.021 & 0.972\\
13.0 & 5.32 & 0.92 $\pm$ 0.02 & 0.745 $\pm$ 0.019 & 0.931\\
\noalign{\smallskip}\hline\hline
\end{tabular}
\end{center}
\label{Statistic}       
\end{table*}


\section{Data analysis}
\label{sec:data}

The \kstarch~is a short-lived particle and its decay vertex cannot be distinguished from the primary collision vertex. It is reconstructed in ALICE via its main decay channel \simplekstarch $\rightarrow$  \kshort + \ppm, which has a branching ratio (B.R.) of (33.3~$\pm$~0.003)\%~\cite{PDG}, taking into account the B.R. of \simplekstarch~$\rightarrow$~\kzero + \ppm~decay and the
probability of \kzero~to be  into a \kshort~state.  
The \kshort~is reconstructed by exploiting its characteristic weak decay topology (\kshort $\rightarrow$ \pion$^+$ + \pion$^-$) into two oppositely charged particles (\VZERO~topology) with branching ratio (69.2~$\pm$~0.05)\%~\cite{PDG}.

\subsection{Pion and \Bkshort~selection}
\label{sec:k0s}

Particle identification for charged pions originating from the primary and secondary vertices (``primary and secondary pions'') is applied on a sample of high-quality tracks reconstructed with the TPC and the ITS. Informations from ITS are required only for primary tracks.  
The primary and secondary tracks reconstructed with the TPC are required to have crossed at least 70 readout rows out of a maximum 159. They are also requested to avoid large gaps in the number of expected tracking points in the radial direction. This is achieved by ensuring that the number of clusters expected, based on the reconstructed trajectory and the measurements in neighbouring TPC pad rows, do not differ by more than 20\%. 
Particles are required to have \pT~$>$~0.15~\gmom~and to be located in the pseudorapidity range \modeta~$<$~0.8 to avoid edge effects in the TPC acceptance. 
Furthermore, tracks of particles possibly originating from weak decays of pions and kaons are rejected when a kink in the track is observed~\cite{ALICE_Performance}.  
Primary tracks are required to be associated with at least one cluster in the SPD and the goodness-of-fit values $\chi^2$~per cluster of the track fit in the ITS and in TPC are restricted in order to select high-quality tracks. 
Primary tracks are required to have a distance of closest approach (DCA) to the primary vertex lower than 2~cm along the beam axis and 7$\sigma$~in the transverse plane, where 
$\sigma$~=~(0.0015+0.0050~\pT$^{-1.1}$)~cm with \pT~in units of \gmom. 
Secondary tracks are required to have a DCA to the primary vertex larger than 0.06~cm. 
Selected pion candidates are identified by requiring that the specific ionization energy loss \dedx~measured in the TPC lies within $n$ standard deviations ($\sigma_{TPC}$) from the specific energy loss expected for pions, with $n$ equal to 3 or 5 for primary and secondary pions, respectively.

The selection criteria used for the \kshort~reconstruction are listed in Tab.~\ref{K0s_cuts}.
Candidates \kshort~are in the rapidity range \modrap~$<$~0.8.  
The distance of closest approach between positively and negatively charged tracks is required to be smaller than one standard deviation with respect to the ideal value of zero and the cosine of the pointing angle ($\theta_{\mathrm{PA}}$), which corresponds to the angle between the \VZERO~momentum and the line connecting the secondary to the primary vertex, is required to be larger than 0.97. 
Only those \VZERO~candidates located at a radial distance larger than 0.5~cm (\VZERO~radius) are used in this analysis. Competing \VZERO~rejection is also applied: the \VZERO~mass is recalculated assuming that one of the pions is a (anti-)proton, and the \VZERO~candidates (about~2\%) are rejected if their mass is compatible with the \rmLambda~mass within $\pm$~0.0043~\gmass, which is about three times the typical mass resolution for the reconstructed $\rmLambda$ in ALICE~\cite{Lambda_pPb}. 
In addition, \kshort~candidates with a proper lifetime larger than 20~cm$/c$ are rejected to remove combinatorial background from interactions with the detector material. The proper lifetime is estimated as \mbox{$Lm_{\textrm{\kshort}}/p$}, where $L$~is the linear (3D) distance between the primary vertex and the \VZERO~decay vertex, $p$~is the total momentum of \kshort, and \mbox{$m_{\textrm{\kshort}}$}~=~0.497611~\gmass~is the nominal \kshort~mass~\cite{PDG}. 
Finally, the invariant mass of \pip\pim~pairs is required to be compatible with the nominal \kshort~rest mass within 
\mbox{$\pm 4\sigma_{m\textrm{\kshort}}$}, with the \kshort~mass resolution value increasing smoothly with the transverse momentum, from $\approx$~\mbox{3.5$\times$10$^{-3}$}~\gmass~at \pT~$\approx$~0 to 
$\approx$~\mbox{6.2$\times$10$^{-3}$}~\gmass~at \pT~=~10~\gmom.  

ALICE has measured \simplekstarZ~exploiting its decay into \K$^{\pm}$ + \pmp~\cite{ALICE_Kstar_5TeV,ALICE_Kstar_8TeV,ALICE_particle_13TeV,Kstar_7TeV,Kstar_PbPb,Kstar_pPb,Kstar_highpt}, with pions and kaons reconstructed as primary particles and identified using energy loss and time-of-flight measurements. The crucial difference in the \simplekstarch~and \simplekstarZ~reconstruction is the charged and neutral kaon identification. In particular, the neutral kaon reconstruction efficiency is larger for \pT~$<$~0.2~\gmom~and for \pT~$>$~2~\gmom.  At low \pT, primary charged kaon detection depends on the tracking efficiency with a threshold of about 0.1~\gmom, whereas at high \pT~the larger efficiency in neutral kaon reconstruction is mainly connected to a loose charged particle selection based on the expected specific energy loss.  

\begin{table*} 
\caption{The selection criteria parameters for \kshort~candidates. DCA stands for distance of closest approach, PV means primary vertex, $\theta_{\mathrm{PA}}$ is the pointing angle, \mbox{$Lm_{\textrm{\kshort}}/p$} is the proper lifetime. The competing \VZERO~rejection window is 1.1157 $\pm$ 0.0043~\gmass~while for the mass of the \pip\pim~pairs the window is
$\left | m_{\textrm{\kshort}} - m_{\textrm{\pip\pim}}  \right | < 4\sigma_{m\textrm{\kshort}}$.}

\begin{center}
\begin{tabular}{ll}
\hline\noalign{\smallskip}
\kshort~selection criteria & Value \\
\hline
\hline\noalign{\smallskip}
Pion \dedx~($\sigma$) & $<$~5 \\ 
DCA of daughter to PV (cm$/c$) & $>$~0.06 \\
DCA between daughters ($\sigma$) & $<$~1 \\
Cosine of $\theta_{\mathrm{PA}}$ & $>$~0.97 \\
\VZERO~radius (cm) & $>$~0.5 \\
Proper lifetime \mbox{$Lm_{\textrm{\kshort}}/p$} (cm) & $<$~20 \\
Competing \VZERO~rejection window (\gmass) & $\pm$0.0043\\
Mass \kshort~window ($\sigma$) & $\pm$4\\
Rapidity \modrap & $<$ 0.8\\
\noalign{\smallskip}\hline\hline
\end{tabular}
\end{center}
\label{K0s_cuts}       
\end{table*}

\begin{figure}[H]
\centering
\includegraphics[width= 0.45\textwidth]{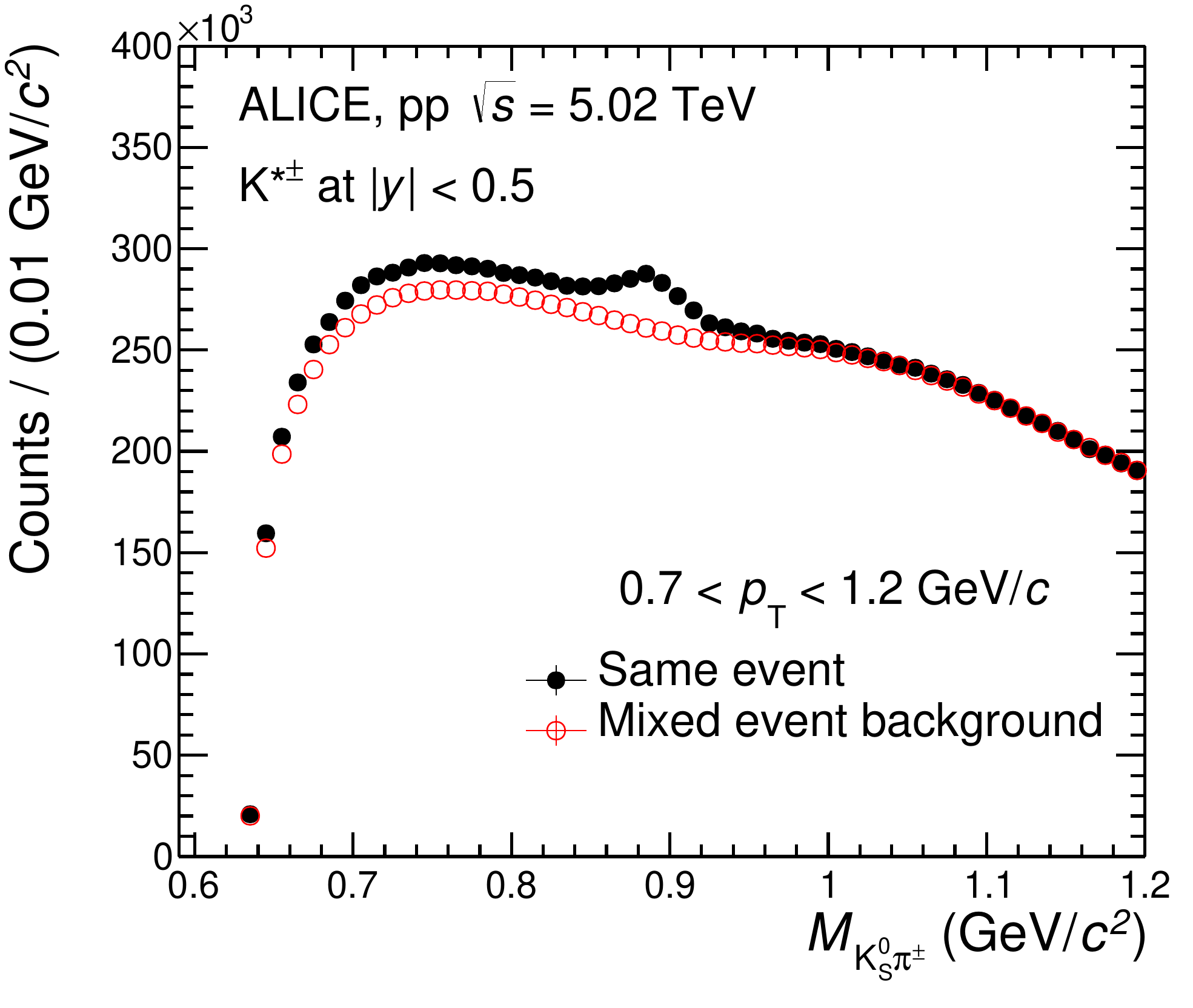}
\includegraphics[width= 0.45\textwidth]{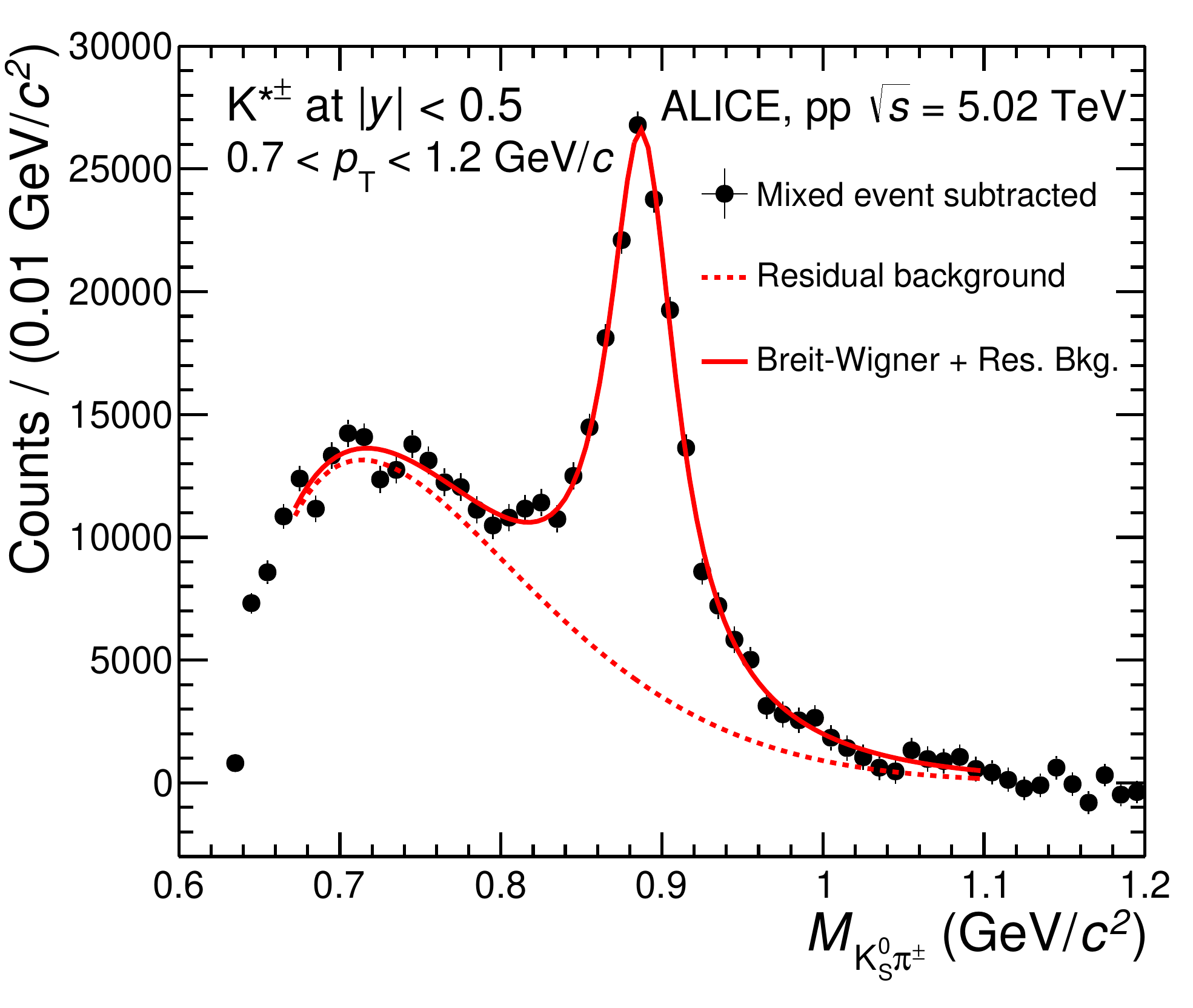}
\includegraphics[width= 0.45\textwidth]{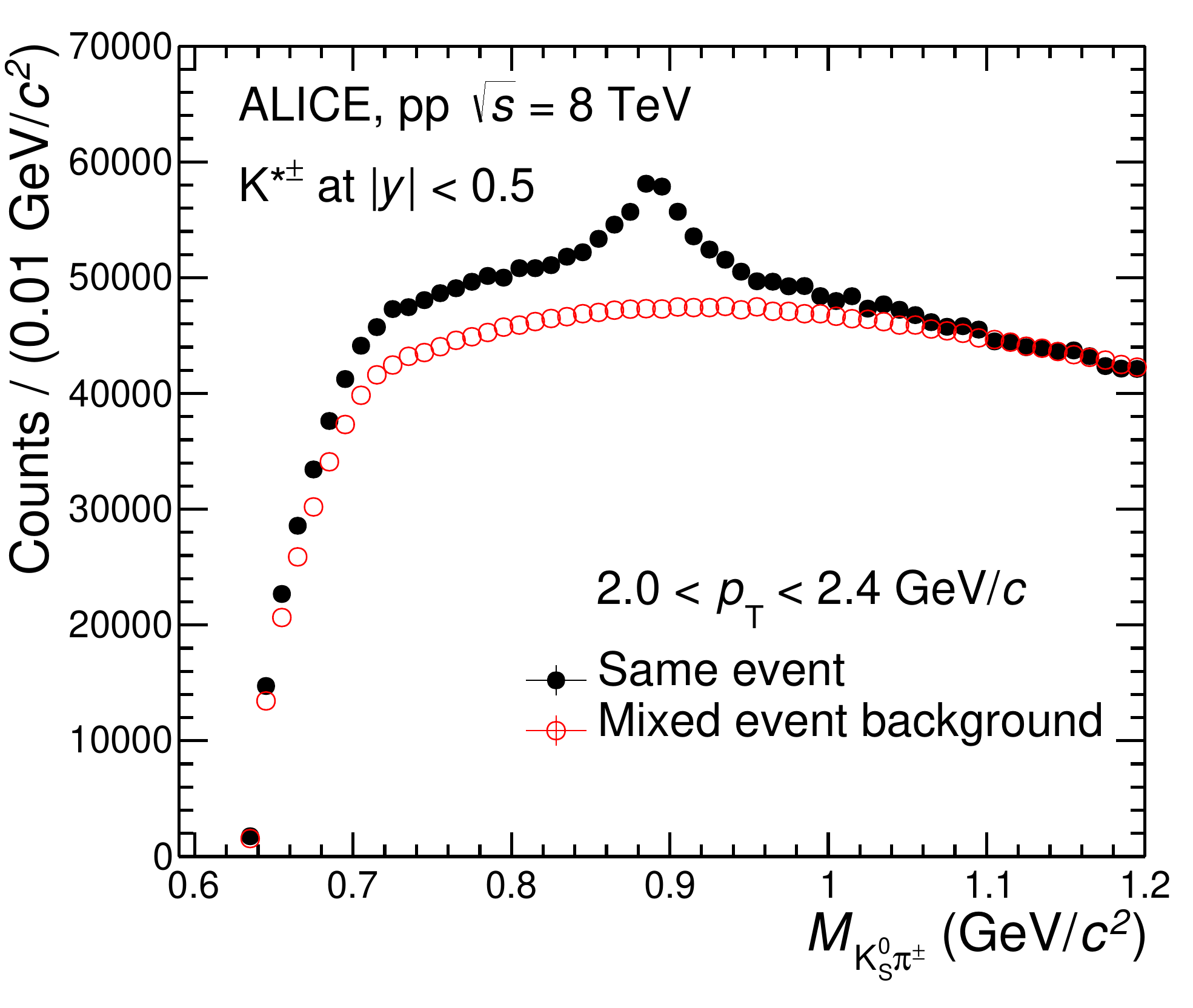}
\includegraphics[width= 0.45\textwidth]{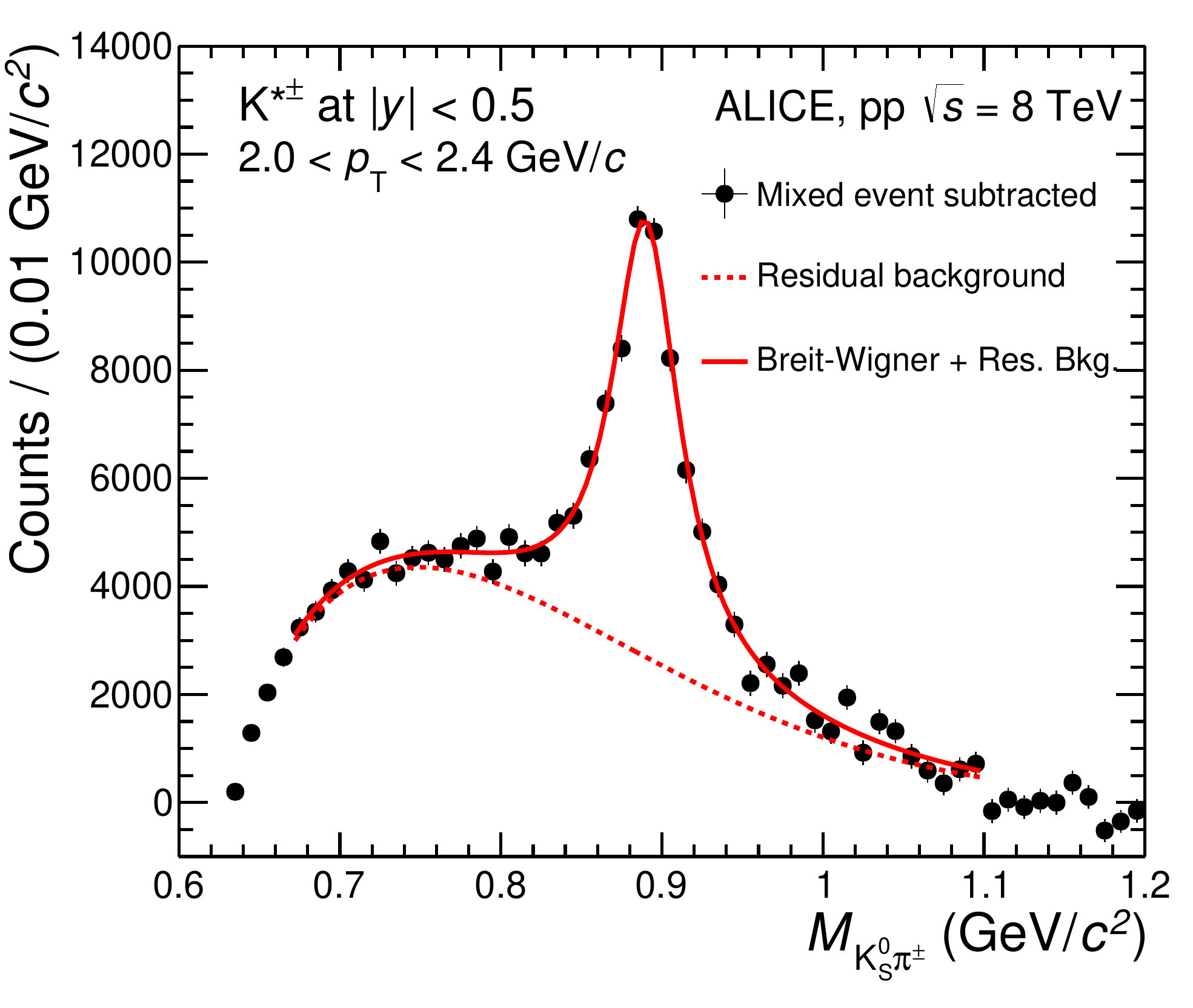}
\includegraphics[width= 0.45\textwidth]{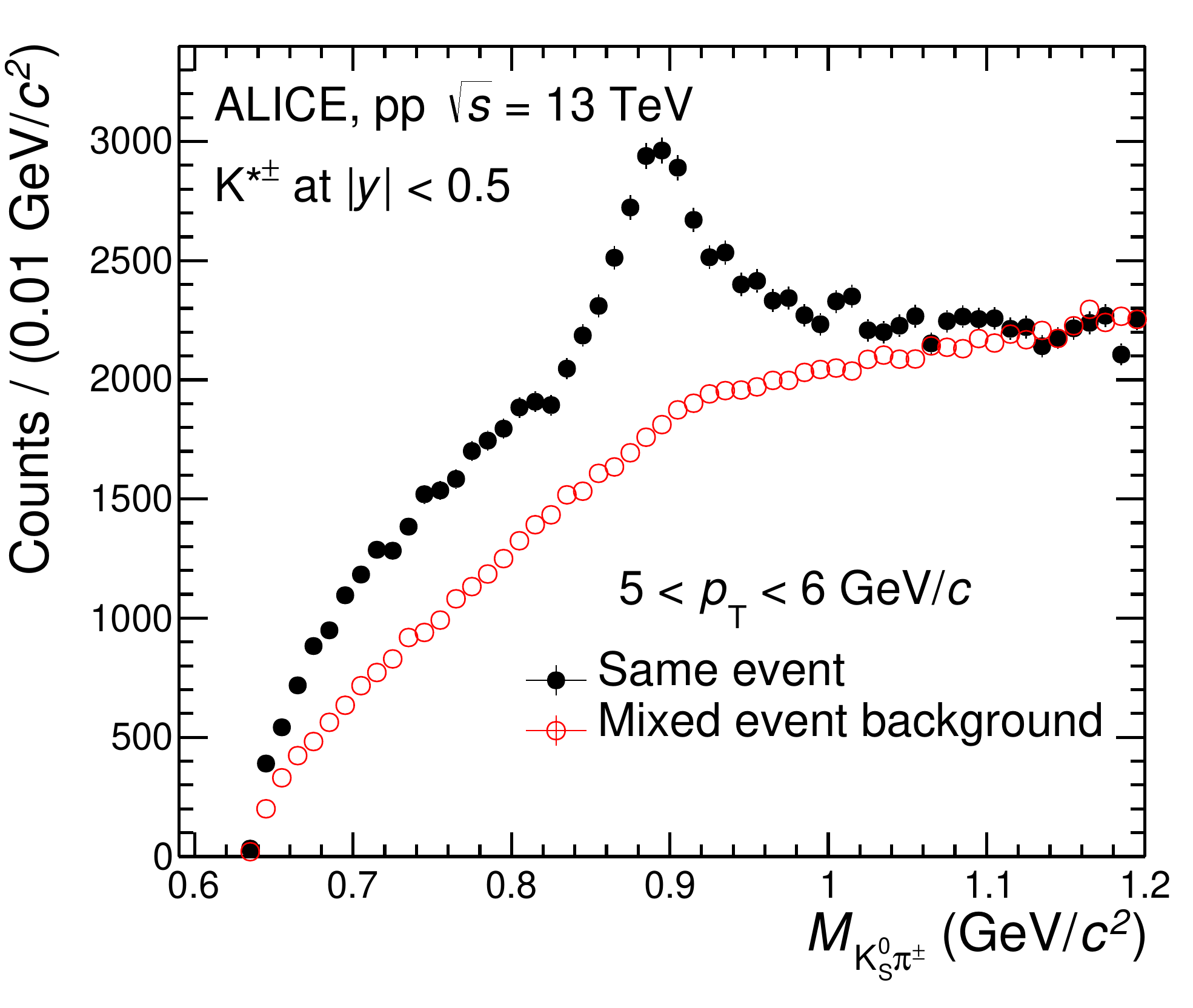}
\includegraphics[width= 0.45\textwidth]{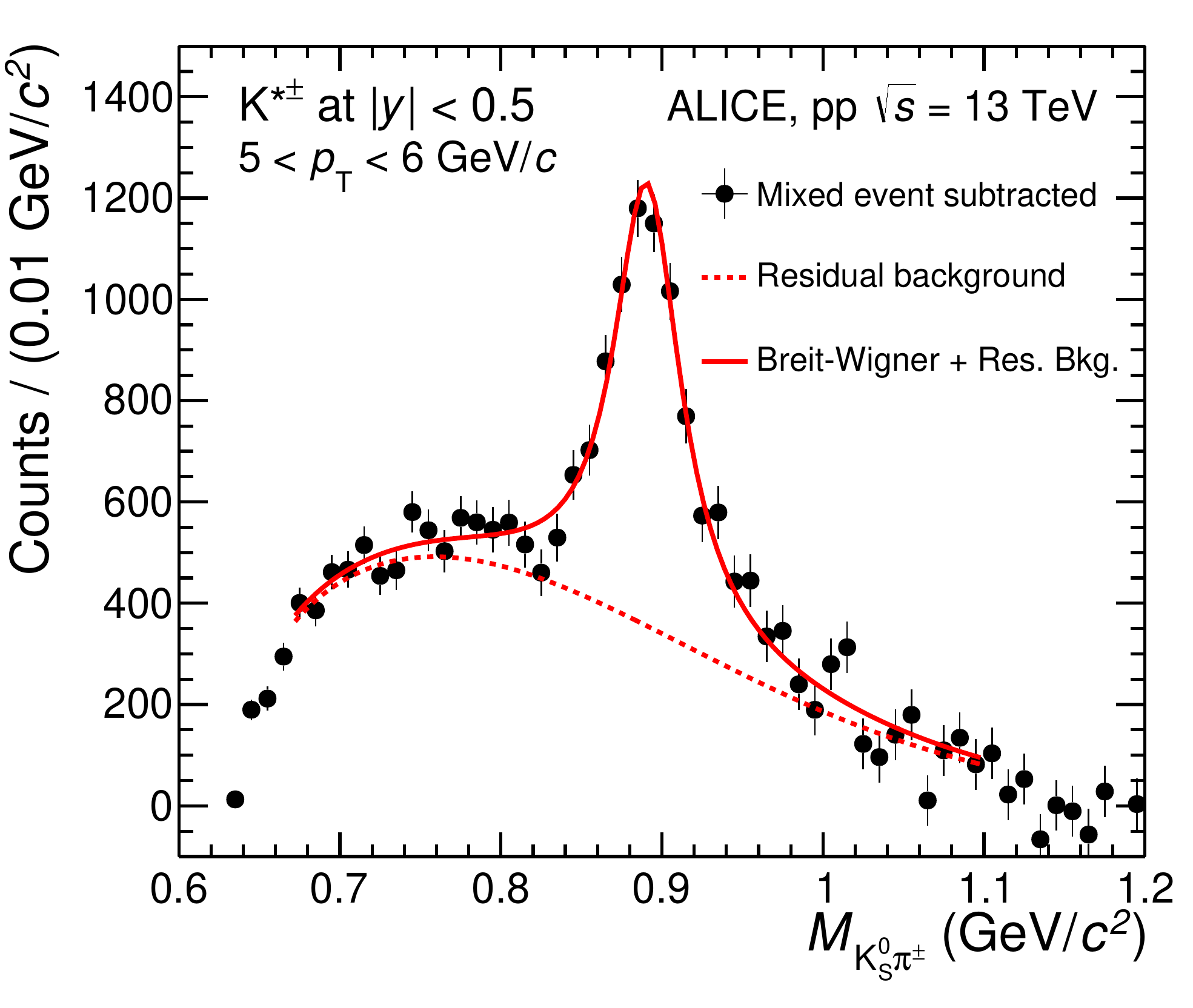}
\caption{(Left panels) The \kshort\ppm~invariant mass distributions at \modrap~$<$~0.5 in pp collisions at \sqrtS~=~5.02, 8, and 13~TeV. The background shape estimated by the event-mixing technique is shown with empty red circles. Statistical uncertainties are shown with error bars.
(Right panels) The \kshort\ppm~invariant mass distributions in pp collisions at \sqrtS~=~5.02, 8, and 13~TeV after background subtraction. The solid red curve is the result of the fit with Eq.~\ref{eqn1}; the dashed red line describes the residual background distribution given by Eq.~\ref{eqn2}. Statistical uncertainties are shown with error bars.}
\label{Fig:signal} 
\end{figure}

\subsection{Signal extraction}
\label{sec:signal}
The raw yield of the \simplekstarch~is extracted from the same-event \kshort\ppm~invariant mass distribution in different \pT~intervals between 0 and 15~\gmom. The nominal mass value~\cite{PDG} is assigned to the \kshort~when the \kshort\ppm~invariant mass is estimated. 
The shape of the uncorrelated background is estimated using the invariant mass distribution of 
\kshort\ppm~pairs selected from different events (event mixing method). To avoid any mismatch due to different acceptances and to ensure a similar event structure, particles from events with similar vertex positions along $z$ ($\Delta z$ $<$ 1~cm) and track multiplicities $n$ ($\Delta n$~$<$~5) are mixed. To reduce statistical uncertainties each event is mixed with 9 others. The mixed-event distribution is then normalized to the same-event distribution in the mass region 1.1~$< M_{\textrm{\kshort\ppm}}<$~1.2~\gmass~and subtracted from the same-event distribution in each \pT~bin. The mixed-event background normalization range is varied for the study of systematic uncertainties.

The \kshort\ppm~invariant mass distributions in different \pT~ranges~obtained for the different collision energies are shown in the left panels of Fig.~\ref{Fig:signal}.
Similar to previous \simplekstarZ~analyses~\cite{ALICE_Kstar_5TeV,ALICE_Kstar_8TeV,ALICE_particle_13TeV,Kstar_7TeV,Kstar_PbPb,Kstar_pPb,Kstar_highpt} the uncorrelated mixed-event background is subtracted from the same-event invariant mass distribution.
The resulting distributions exhibit a characteristic peak on top of a residual
background, as reported in the right panels of Fig.~\ref{Fig:signal}. 
The latter is due to the presence of correlated pairs from jets, multi-body decays of heavier particles and misreconstructed resonance decays. 
The resulting distribution is fitted with a combination of the non-relativistic Breit-Wigner function to describe the signal peak and a $F_{BG}$ function to describe the residual background. 

The fit, based on the minimization of the $\chi^{\textrm{2}}$, was performed according to the following expression:

\begin{eqnarray}
\frac{dN} {dM_{\textrm{\kshort\ppm}}} = \frac{C}{2\pi}\frac{\Gamma_{\textrm{0}}}{\left(M_{\textrm{\kshort\ppm}}- M_{\textrm{0}}\right)^{2}+  \frac{\Gamma_{\textrm{0}}^{2}}{4}} + F_{BG}\left(M_{\textrm{\kshort\ppm}}\right)
\label{eqn1}
\end{eqnarray}
\noindent where $M_{\textrm{0}}$ and $\Gamma_{\textrm{0}}$ are the mass and the width of the \simplekstarch~\cite{PDG}. 
 The $C$ parameter is the integral of the peak function from 0 to $\infty$. The detector mass resolution for the reconstruction of \simplekstarch~is negligible compared to its natural width, $\Gamma_{\textrm{0}}$~=~(0.0508~$\pm$~0.0009)~\gmass~\cite{PDG}, and it is therefore not included in the peak model.
The mass and width of \simplekstarch~were found to be compatible with the values reported in~\cite{PDG}. For the measurement of the yields, the width of \simplekstarch~was fixed to its natural value. Fits were performed with the width kept as a free parameter or fixed at 0.0517 or 0.0499 \gmass~to estimate the systematic uncertainty.

The shape of the correlated background in the invariant mass distribution of \kshort\ppm~pairs is studied using  the same samples of simulated events described in Sect.~\ref{sec:efficiency} that were used to  estimate the Acceptance$\times$Efficiency corrections. The produced particles and their decay products are propagated through the ALICE detector using GEANT3~\cite{GEANT}. Invariant mass distributions for \kshort\pion$^+$~and  \kshort\pion$^-$~pairs are accumulated after applying  the same event, track and particle identification selections as in data. The study shows that after subtracting the combinatorial background, the remaining background has a smooth dependence on mass. It is well described by the following function, already used in Refs.~\cite{Kstar_LEP,OPAL_background}:

\begin{eqnarray}
F_{BG}\left(M_{\textrm{\kshort\ppm}}\right) = \left[M_{\textrm{\kshort\ppm}} - \left(m_{\textrm{\ppm}} + m_{\textrm{\kshort}}\right)\right]^n 
\exp \left(a + b M_{\textrm{\kshort\ppm}} + c M_{\textrm{\kshort\ppm}}^2 \right)
\label{eqn2}
\end{eqnarray}
\noindent where $n$, $a$, $b$, and $c$ are fit parameters and $m_{\textrm{\ppm}}$ and $m_{\textrm{\kshort}}$ are the pion and \kshort~masses~\cite{PDG}. 
Examples of these fits for different \pT~intervals and different pp collision energies are shown in the right panels of Fig.~\ref{Fig:signal}. The typical fitting interval was \mbox{0.66 $< M_{\textrm{\kshort\ppm}} <$ 1.1~\gmass}.

The \simplekstarch~raw yield ($N_{\mathrm{raw}}$) is determined by integrating the combinatorial background-subtracted invariant mass distribution over the interval 0.79$-$0.99~\gmass, subtracting the integral of the residual background fit function over the same range, and correcting the result to account for the yield outside that range. The yield in the tails is estimated by integrating the non-relativistic Breit-Wigner function from \mbox{$m_{\textrm{\ppm}} + m_{\textrm{\kshort}}$} to 0.79~\gmass~and from 0.99~\gmass~to infinity.
This correction to the total yield is about 13\%. As an alternative used to estimate the systematic uncertainties, the \simplekstarch~yield is also obtained by integrating the peak fitting function in the allowed region (\mbox{$m_{\textrm{\ppm}} + m_{\textrm{\kshort}}, \infty$}).  

\begin{figure*}[h]
\centering
\includegraphics[width= 0.60\textwidth]{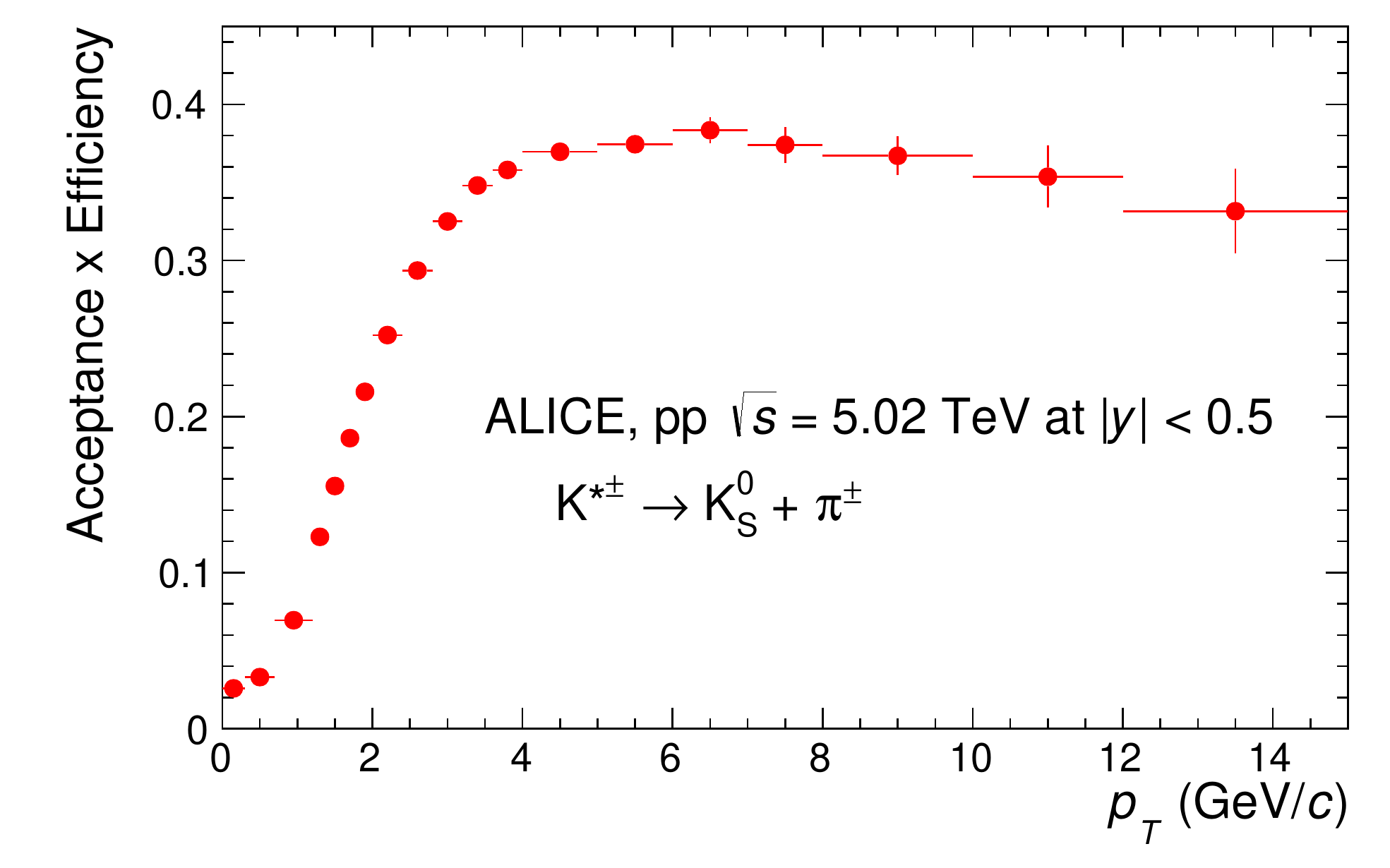}
\caption{Acceptance$\times$Efficiency as a function of \pT~for \simplekstarch~mesons, detected by their decay to
\kshort~+~\ppm, with \kshort~reconstructed by their decay to \pip~+~\pim. The \kshort~$\rightarrow$~\pip~+~\pim branching ratio is included in the efficiency estimation. Statistical uncertainties are shown with error bars.}
\label{Efficiency} 
\end{figure*}

\subsection{Efficiency and acceptance}
\label{sec:efficiency}
To obtain the corrected resonance yields, the convolution between the geometrical acceptance ($A$) and the resonance reconstruction efficiency ($\epsilon_{\mathrm{rec}}$), which takes into account the criteria used to select primary charged pions and \kshort, is determined. 
The $A\times\epsilon_{\mathrm{rec}}$ product takes into account also the branching ratio of 
\kshort~$\rightarrow$~\pip~+~\pim. For each collision energy, 
$A\times\epsilon_{\mathrm{rec}}$ is determined using samples of about 50~million pp events simulated with different Monte Carlo event generators (PYTHIA6-Perugia 2011 tune~\cite{PYTHIA6,Perugia2011}, PYTHIA8-Monash 2013 tune~\cite{PYTHIA8,Monash2013}, 
EPOS-LHC~\cite{EPOS}) and a GEANT3-based simulation~\cite{GEANT} of the ALICE detector response. The actual positions of the detectors (alignment), maps of dead or noisy elements, and time and amplitude calibrations are used in the reconstruction of real and simulated data. All the parameters taken into account for a careful calibration of the ALICE detector are listed in~\cite{ALICE_Performance}.  The residual differences between data and the sample of Monte Carlo simulation previously described are considered in the systematic uncertainty.  
  
For each \pT~interval, the $A\times\epsilon_{\mathrm{rec}}$ is calculated as the ratio $N_{\mathrm{rec}}/N_{\mathrm{gen}}$, where $N_{\mathrm{rec}}$ is the number of particles reconstructed in the \kshort~+~\ppm~channel after all event
and particle selections, while $N_{\mathrm{gen}}$ is the number of generated mesons decaying in the same channel. Both generated and reconstructed mesons have the rapidity in the range \modrap~$<$~0.5.  
In general, the efficiency depends on the shape of the generated particle \pT~spectrum. Therefore, at the different collision energies, the efficiency for \simplekstarch~is estimated re-weighting iteratively  the shape of the generated \pT~spectrum to measured shape.
As an example the transverse momentum dependence of $A\times\epsilon_{\mathrm{rec}}$ is reported in Fig.~\ref{Efficiency} for the \sqrtS~=~5.02~TeV sample.

\subsection{Yield corrections}
\label{sec:yield}

The differential transverse momentum yield for inelastic pp collisions was calculated as

\begin{eqnarray}
\frac{1}{N_{\mathrm{INEL}}}\frac{d^{2}N}{dp_{T}dy} = \frac{N_{\mathrm{raw}}}{N_{\mathrm{MB}} \times {\mathrm{B.R.}} \times \Delta \pT \times \Delta y}
\frac{f_{\mathrm{SL}}}{(A\times\epsilon_{\mathrm{rec}})}\times\epsilon_{\mathrm{trig}}\times\epsilon_{\mathrm{vertex}} .
\end{eqnarray}

The raw yields are corrected for the resonance branching ratio (B.R.~=~33.3$\%$) and $A\times\epsilon_{\mathrm{rec}}$ in the \kshort~+~\ppm~channel.
Furthermore, these yields were normalized to the number of minimum bias events $N_{\mathrm{MB}}$ and corrected for the vertex reconstruction efficiency 
$\epsilon_{\mathrm{vertex}}$ as well as for the trigger selection efficiency $\epsilon_{\mathrm{trig}}$. 
Values of $N_{\mathrm{MB}}$, $\epsilon_{\mathrm{vertex}}$, and $\epsilon_{\mathrm{trig}}$ for all collision energies are reported in Tab.~\ref{Statistic}. 
The signal-loss correction $f_{\mathrm{SL}}$ takes into account the fraction of \simplekstarch~mesons in non-triggered inelastic events and it is estimated by Monte Carlo simulations.
The latter is a \pT-dependent correction factor which has its maximum at low \pT~($f_{\mathrm{SL}}\approx$~1.04 for \pT~$<$~1~\gmom~and $f_{\mathrm{SL}}~\approx$~1.01 for 
\pT~$>$~1~\gmom).

\subsection{Systematic uncertainties}
\label{sec:systematics}

The measurement of \simplekstarch~production in pp collisions was tested for systematic effects  
due to uncertainties in signal extraction, track selection criteria and particle identification for primary pions, \kshort~reconstruction, global tracking efficiency for primary pions, primary vertex selection window, knowledge of the ALICE material budget and hadronic interaction cross section used in simulations and signal loss correction, as summarized in Tab.~\ref{syst}. 
The yield-weighted mean values are quoted for three separate transverse momentum intervals: low (0~$<$~\pT~$<$~1.2~\gmom), intermediate (1.2~$<$~\pT~$<$~4~\gmom), and high-\pT~(4~$<$~\pT~$<$~15~\gmom).

The systematic uncertainties are dominated by the raw yield extraction, labeled as ``Signal extraction" in Tab.~\ref{syst} and amount to about 3-6\%. This includes the sensitivity in the choice of the normalization interval, the fitting range, the shape of the residual background function, the bin counting range and the constraints on the resonance width imposed in the fitting procedure.
In addition to the default strategy described in Sec~\ref{sec:signal}, the combinatorial background was normalized in different invariant mass regions. The sensitivity of the \simplekstarch~yield extraction to the fit range was studied by varying each interval boundary by  
$\pm$~0.005~\gmass. As an alternative to the function used to describe the shape of the residual background (Eq.\ref{eqn2}), a third- and a second-order polynomial function was used. In this last case, the fitting range was restricted to the region 0.74-1.1~\gmass, where the background is reasonably approximated by a second order polynomial shape.  The integration limits were varied by $\pm$~0.01~\gmass. The sensitivity of the fit to the constraint on the \simplekstarch~signal width was estimated by using width values that take into account the current uncertainty on the PDG average value (0.0009~\gmass~\cite{PDG}) or by fitting the signal without any constraint. 

The contribution to the uncertainty related to the primary charged pion reconstruction, reported in Tab.~\ref{syst}, was estimated by varying simultaneously in the data and Monte Carlo events the track and the PID selections. This uncertainty ranges from 1 to 2$\%$. In particular, the sensitivity of the track selection on the number of crossed rows, the number of reconstructed TPC space points and the distance of closest approach to the primary vertex was tested. To study the effect of PID on the signal extraction, the selection criteria based on the TPC energy loss were varied with respect to the default setting described in Sec.~\ref{sec:k0s}. PID criteria of 2.5$\sigma_{TPC}$ and 4$\sigma_{TPC}$ were used.  

Systematic uncertainties due to the \VZERO~topological and \kshort~secondary track selections are reported in Tab.~\ref{syst} under label ``\kshort~reconstruction". These uncertainties were estimated by varying simultaneously in the data and Monte Carlo events the track and the PID selection criteria for the secondary tracks, and by varying all the topological selection criteria (DCA of decay products to PV and between decay products, cosine of pointing angle and \VZERO~radius).
The sensitivity of the measurement to the competing \VZERO~rejection, the mass selection, the \kshort~rapidity range and lifetime was also studied by varying the interval selections. Relative uncertainties in the range 0.7-2.9$\%$ were estimated for the three energies in all the \pT~intervals. The total systematic uncertainties associated with the \kshort~measurement are lower than those for the charged ones~\cite{ALICE_7TeV_multiplicity,ALICE_particle_13TeV}. In particular, by exploiting the topological identification of \kshort, the large uncertainties (amounting to about~6$\%$) originating from track selection and the PID procedure for \K$^{\pm}$~are avoided. 

In ALICE, the track reconstruction proceeds from the outermost to the innermost radius of the TPC. To have a high-quality track for a particle originating from the primary vertex, the segment of track reconstructed in the TPC should be matched to reconstructed points in the ITS. This is not necessary for secondary tracks that originate from weak decay vertices.  
The differences in matching probabilities of TPC tracks with reconstructed points in the ITS between data and Monte Carlo simulations define the global tracking efficiency uncertainty.
This uncertainty is in the range 1-1.4$\%$ for the 5.02~TeV data set, while a constant value of 1$\%$ and 3$\%$ was estimated for the 13 and 8~TeV data, respectively. These uncertainties are correlated across
 \pT~for the inspected data sets.
Variations in the selection window around the primary vertex position can modify the yield by about 0.6-2$\%$.  
The uncertainty related to the knowledge of the ALICE material budget ranges from 3.1$\%$~to 1.7$\%$ for \mbox{\pT~$<$~4~\gmom}
and is about 0.7$\%$ at higher \pT. The uncertainty connected to the knowledge of the hadronic interaction cross section in the detector material is about 1$\%$ for \pT~$<$~4~\gmom. These effects are evaluated combining  the uncertainties for a $\pi$~and a \kshort, determined as in~\cite{ALICE_particle_13TeV,ALICE_multistrange_13TeV}, according to the kinematics of the decay. For the signal loss correction an uncertainty of about 1.5$\%$ was estimated for 
\pT~$<$~1.2~\gmom~for 5.02 and 13~TeV collisions, while a slightly lower value was estimated for the 8 TeV collisions. This, for each \pT~interval, is the largest value between one half of ($f_{\mathrm{SL}}$~-~1) and the difference of signal-loss correction values estimated with different event generators.

The total systematic uncertainty is 4 - 8$\%$ for all the considered \pT~intervals  
whereas the systematic uncertainties assigned to the \simplekstarZ~measurements performed to date range from 9$\%$ to 18$\%$ depending on energy and \pT~\cite{ALICE_Kstar_5TeV,ALICE_Kstar_8TeV,ALICE_particle_13TeV}. This confirms that the systematic uncertainty on the \K$^*/$\K~ratio can be reduced by studying the charged resonant state.

\begin{table*}
\caption{Sources and yield-weighted mean values of the relative systematic uncertainties (expressed in $\%$) on the differential yields of the \simplekstarch~resonance at the three centre-of-mass energies under study for low, intermediate and high-\pT~ranges. }
\begin{center}
\resizebox{\textwidth}{!}{
\begin{tabular}{lccc}
\hline\noalign{\smallskip}
\sqrtS~(TeV)~~~~~~~~~~~~~~~~~~~~~~~~~~~~~~~~~~~~~&~~~~~~~~~5.02~~~~~~~~~~~~~~~&~~~~~~~~~~~~~8.0~~~~~~~~~~~~&~~~~~~~~~~~~~~~~13.0~~~~~~~~~~~~~~ \\
\noalign{\smallskip}\hline
\hline
\end{tabular}}
\resizebox{\textwidth}{!}{
\begin{tabular}{lccccccccc}
\pT~(\gmom) & 0 -- 1.2 & 1.2 -- 4 & 4 -- 15 & 0 -- 1.2 & 1.2 -- 4 & 4 -- 15 & 0 -- 1.2 & 1.2 -- 4 &4 -- 15 \\
\hline\noalign{\smallskip}
Signal extraction ($\%$) & 5.4 & 2.8 & 3.4 & 5.8 & 5.5 & 5.4 & 4.4 & 3.7 & 4.5 \\
Primary pion reconstruction ($\%$) & 1.2  & 1.0 & 1.0 & 1.2 & 1.1 & 1.5 & 2.1 & 1.4 & 1.3 \\
\kshort~reconstruction ($\%$) & 0.8  & 0.7 & 1.0 & 2.9 & 0.9 & 0.9 & 2.2 & 1.3 & 1.2 \\
Global tracking efficiency ($\%$) & 1.0  & 1.0 & 1.4 & 3.0 & 3.0 & 3.0 & 1.0 & 1.0 & 1.0 \\
Primary vertex ($\%$) & 2.3 & 0.7 & 1.4 & 1.5 & 0.6 & 1.5 & 1.0 & 0.6 & 0.7 \\
Material budget ($\%$) & 3.1  & 1.7 & 0.7 & 3.1 & 1.7 & 0.7 & 3.0 & 1.6 & 0.7 \\
Hadronic interaction ($\%$) & 1.1 & 1.1 & 0.5 & 1.1 & 1.1 & 0.5 & 1.1 & 1.1 & 0.5 \\
Signal Loss ($\%$) & 1.4  & 0.6 & 0.4 & 0.9 & 0.4 & 0.1 & 1.6 & 0.7 & 0.5 \\
\hline\noalign{\smallskip}
Total ($\%$) & 7.1  & 3.9 & 4.3 & 8.1 & 6.8 & 6.6 & 6.6 & 4.8 & 5.1 \\

\noalign{\smallskip}\hline\hline
\end{tabular}}
\end{center}
\label{syst}       
\end{table*}

\begin{figure*}[h]
\centering
\includegraphics[width= 0.60\textwidth]{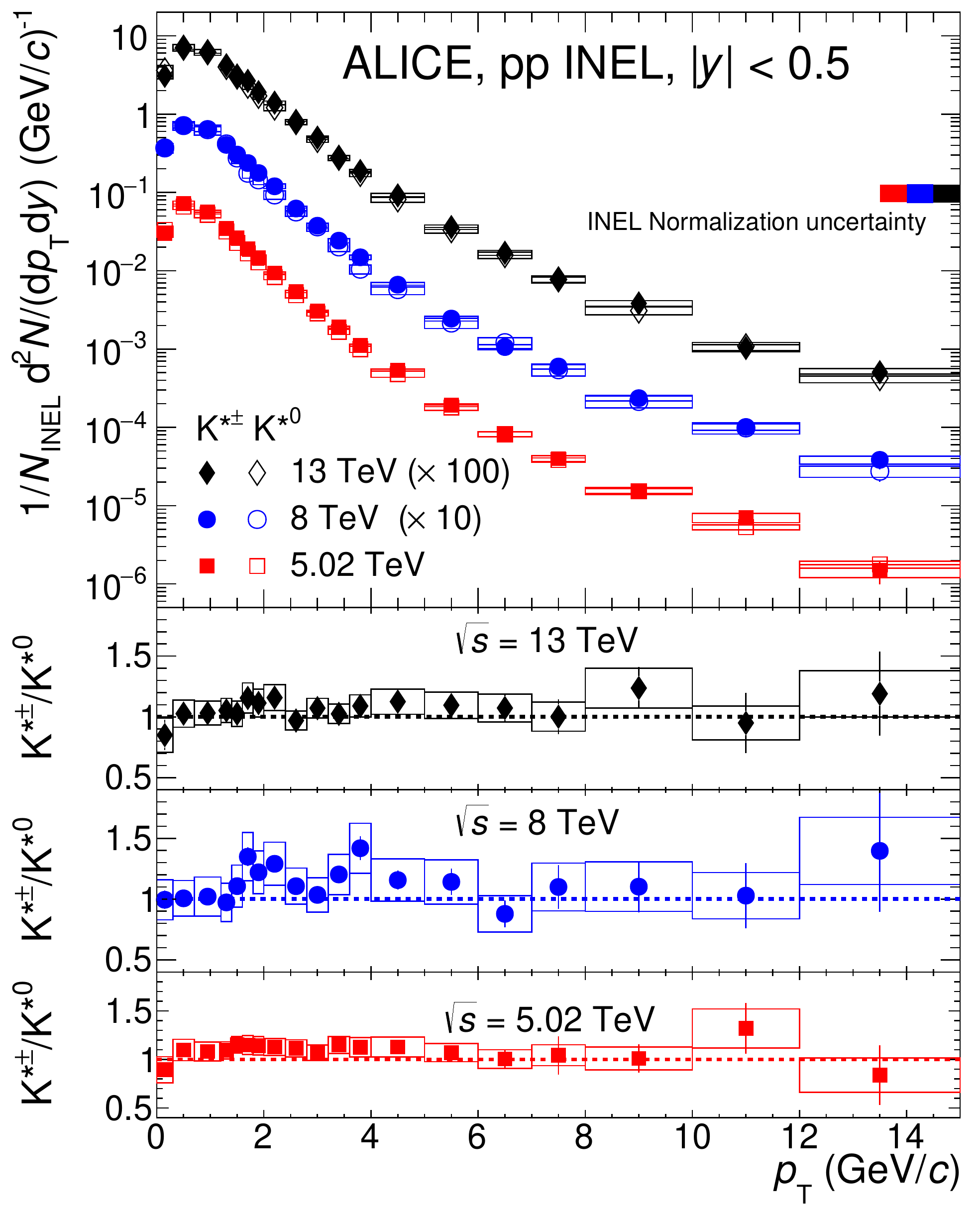}
\caption{(Colour online) The \pT~spectra of \simplekstarch~in inelastic pp collisions at \sqrtS~=~5.02, 8, and 13~TeV (full symbols) are compared to the \pT~spectra~of \simplekstarZ~mesons (open symbols) at the same 
energies~\cite{ALICE_Kstar_5TeV,ALICE_Kstar_8TeV,ALICE_particle_13TeV}. Statistical and systematic uncertainties are reported as error bars and boxes, respectively. The normalization uncertainties (2.51\%, 2.72\%, and 2.55\% for 5.02, 8, and 13~TeV, respectively, see Tab.~\ref{Statistic}) are indicated as coloured boxes and are not included in the point-to-point uncertainties. The ratio of each measured 
\pT~distribution for \simplekstarch~mesons at \sqrtS~=~5.02 (red points), 8 (blue points)  and 13~TeV (black points) to the \simplekstarZ~spectrum at the same collision energy is reported in the bottom panels. The systematic uncertainty due to global tracking,
material budget and hadronic interaction cross section of primary pions are equal for charged and neutral \rsimplekstar, thus they cancel out in the propagation of the uncertainty to the final ratio.}
\label{Kstar_K0_spectra} 
\end{figure*}

\begin{figure*}[h]
\centering
\includegraphics[width= 1.0\textwidth]{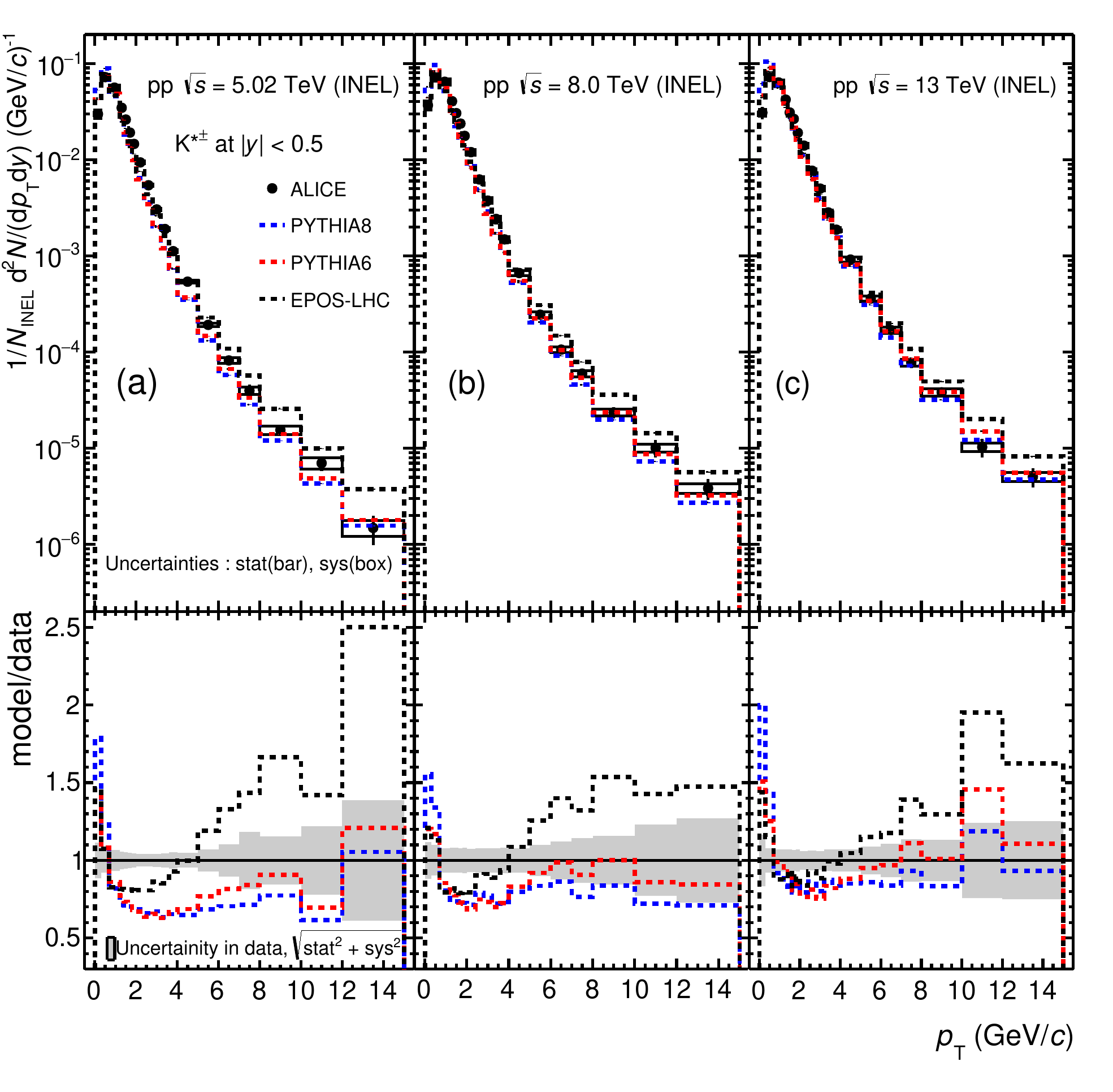}
\caption{(Colour online) The \simplekstarch~\pT~spectra (black dots) measured in inelastic pp collisions at (a)\sqrtS~=~5.02~TeV, (b) 8~TeV, and (c) 13~TeV are compared to the distributions predicted by PYTHIA8-Monash 2013~\cite{Monash2013} (blue lines), PYTHIA6-Perugia 2011~\cite{Perugia2011} (red lines), and EPOS-LHC~\cite{EPOS} (black lines). Statistical and systematic uncertainties are shown with error bars and empty boxes, respectively. The ratios of the rebinned predictions to the measured distributions are reported in the bottom panels. The shaded bands represent the fractional uncertainties of the data points.}
\label{Model_comparison_total} 
\end{figure*}

\begin{figure*}[h]
\centering
\includegraphics[width= 0.45\textwidth]{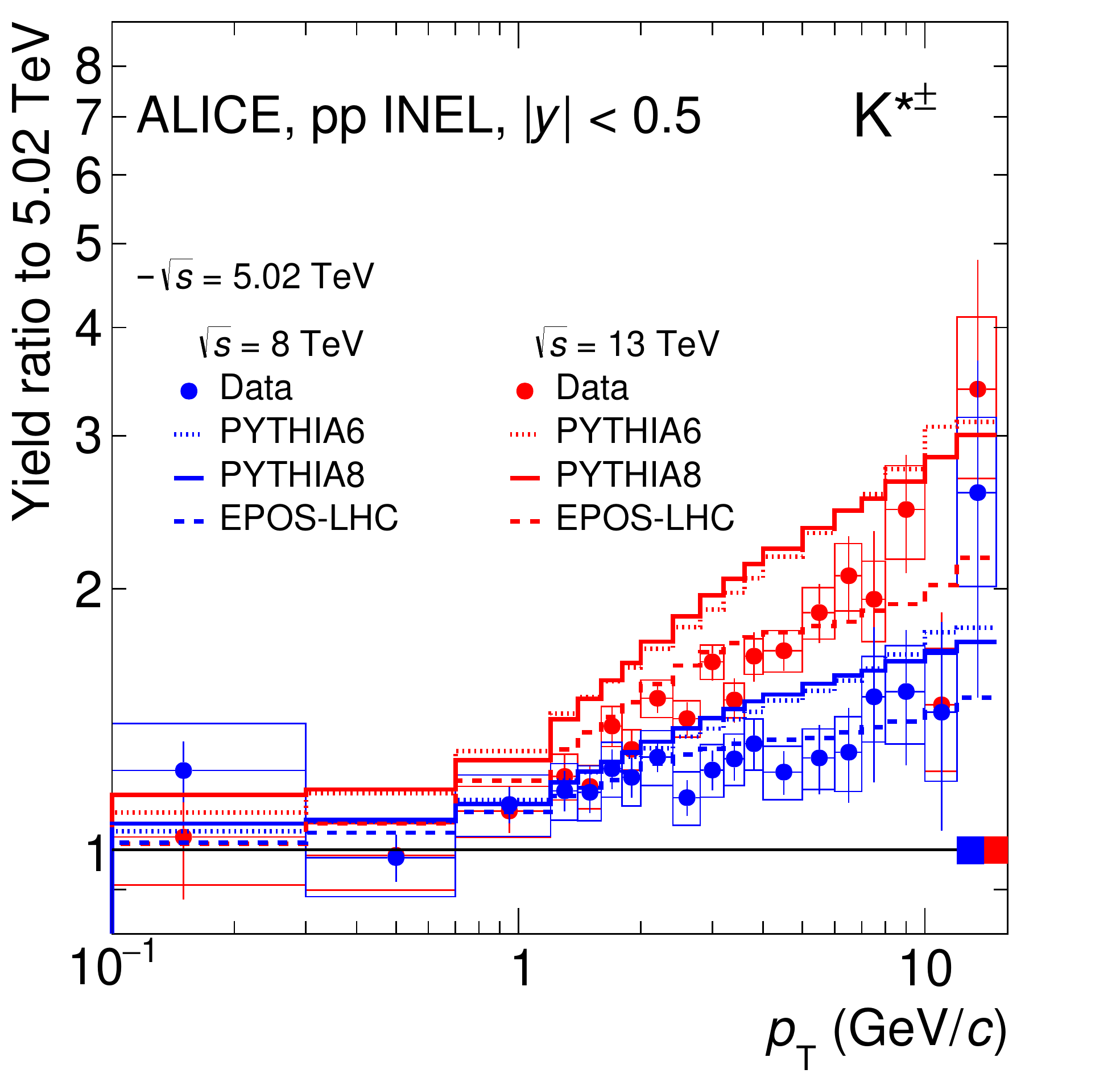}
\includegraphics[width= 0.45\textwidth]{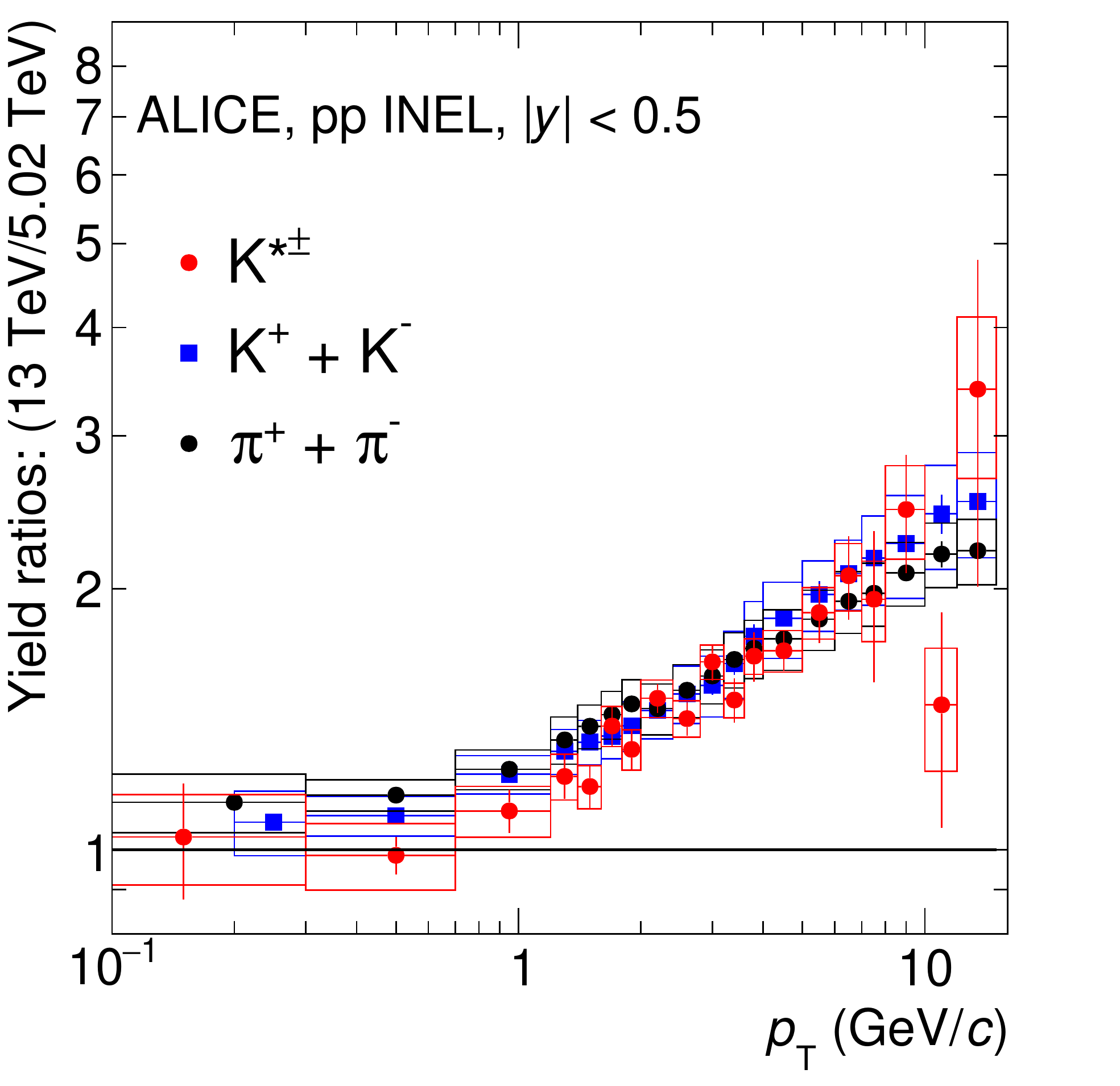}
\caption{(Left panel) Ratios of transverse momentum spectra of \simplekstarch~in inelastic pp events at \sqrtS~=~8 and 13~TeV to corresponding spectra at 5.02~TeV. Statistical and systematic uncertainties are shown with error bars and empty boxes, respectively. The normalization uncertainties are shown as coloured boxes around 1 and they are not included in the point-to-point uncertainties. 
Blue and red 
histograms represent the predictions for the same ratios from PYTHIA6 Perugia 2011, PYTHIA8 Monash 2013, and EPOS-LHC.
(Right panel) Ratios of transverse momentum spectra of \simplekstarch, $\Kp +\Km$~and 
$\pionp + \pionm$~in inelastic pp events at \sqrtS~=~13~TeV~\cite{ALICE_particle_13TeV} to corresponding spectra at 5.02~TeV~\cite{ALICE_particle_5TeV}. Statistical and systematic uncertainties are shown with error bars and empty boxes, respectively.}
\label{Energy_dependence} 
\end{figure*}

\begin{figure*}[h]
\centering
\includegraphics[width= 0.80\textwidth]{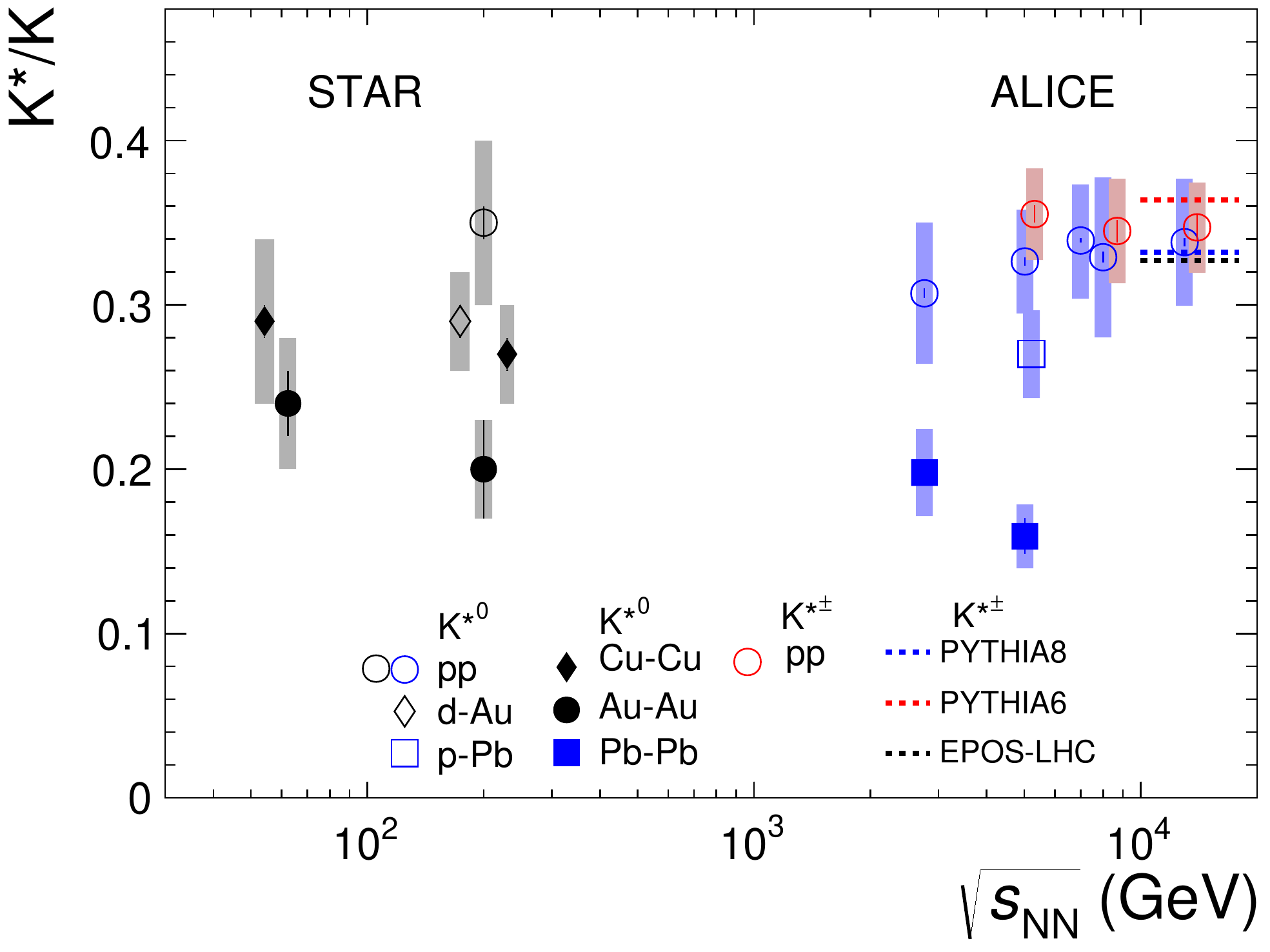}
\caption{(Colour online) Particle ratios \simplekstarch/K~and \simplekstarZ/K, depicted as \rsimplekstar/K, in pp~\cite{Kstar_PbPb,Kstar_7TeV,Kstar_highpt,ALICE_particle_7TeV,ALICE_Kstar_5TeV,
ALICE_Kstar_8TeV,ALICE_particle_13TeV,ALICE_particle_5TeV,
STAR_Kstar_200,STAR_Kstar_CuCu_AuAu}, central \mbox{d--Au}~\cite{STAR_Kstar_dAu}, central \mbox{p--Pb}~\cite{Kstar_pPb}  and central 
\mbox{A--A}~\cite{Kstar_PbPb,Kstar_highpt,ALICE_particle_276Pb,STAR_Kstar_200,STAR_Kstar_CuCu_AuAu} collisions as a function of \sqrtSnn. For the \mbox{d--Au} data, the numerator yield is derived from a combination of \simplekstarZ~and \simplekstarch~states. Bars represent the statistical uncertainties and boxes represent the systematic uncertainties. The points for \simplekstarZ~for \mbox{d--Au},
\mbox{Cu--Cu}~and \mbox{p--Pb} collisions and for \simplekstarch~for pp collisions have been shifted horizontally for visibility. Red, blue and black lines represent the \simplekstarch/K ratio predicted with PYTHIA6-Perugia 2011~\cite{Perugia2011},  PYTHIA8-Monash 2013~\cite{Monash2013} and EPOS-LHC~\cite{EPOS}, respectively.} 
\label{Kstar_K_vs_energy} 
\end{figure*}

\section{Results and discussion}
\label{sec:results}

\subsection{Energy dependence of \BpT~spectra and model comparison}
\label{sec:dependence}
The first measurement of \simplekstarch~meson production in inelastic pp collisions at \sqrtS~$=$~5.02, 8, and 13~TeV up to \pT~=~15~\gmom~is presented in Fig.~\ref{Kstar_K0_spectra}. The \pT-differential yields of \simplekstarch~are compared to those previously measured for \simplekstarZ~in the same collision systems~\cite{ALICE_Kstar_5TeV,ALICE_Kstar_8TeV,ALICE_particle_13TeV}.
The spectra of the charged and neutral mesons are consistent within the uncertainties, as expected considering the similarity of their quark content and mass.

A comparison between the measured \pT~spectra and predictions based on QCD-inspired event generators  such as PYTHIA6~\cite{PYTHIA6}, PYTHIA8~\cite{PYTHIA8} and EPOS-LHC~\cite{EPOS} provides useful information on the hadron production mechanisms.

Event generators such as PYTHIA combine a perturbative formalism of hard processes with a non-perturbative description of hadronization that is simulated using the Lund string fragmentation model~\cite{Lund}. In the PYTHIA tunes considered here, multiple parton-parton interactions in the same event and the colour reconnection mechanism are taken into account. These effects are important in hadron-hadron interactions at the high LHC energies. In particular, colour string formation between final-state partons may mimic effects similar to those induced by collective flow in heavy-ion collisions~\cite{CR}.

The PYTHIA6-Perugia 2011 tune takes into account some of the lessons learnt from the early LHC data from inelastic pp collisions at 0.9 and 7 TeV. For instance, it takes into account the observed increase in baryon production in the strangeness sector by tuning 
the $\Lambda/\K$~ratio on the ALICE~\cite{Strangeness_900GeV,Particle_900GeV} and CMS~\cite{CMS_strangeness} data. On the other hand, the 
\simplekstarZ/\K~ratio is tuned on the LEP measurements~\cite{Perugia2011}. Monash 2013 is an updated set of parameters for the PYTHIA8 event generator, with particular attention to heavy-quark fragmentation and strangeness production.
For all studied LHC collision energies the PYTHIA predictions overestimate by a factor of 1.5\mbox{--}2 
the \simplekstarZ~production at transverse momenta below 0.5~\gmom~and underestimate its production by about 10-20\%~at \pT~$>$~1~\gmom~\cite{ALICE_Kstar_8TeV,ALICE_particle_13TeV,Kstar_7TeV}.

The EPOS-LHC event generator differs significantly from PYTHIA in its modeling of both the hadronization and the underlying event. It is a microscopic model that relies on parton-based Gribov-Regge theory  
with an improved flow parameterization  which takes into account the case of a very dense system in a small volume. This high density core is produced by the overlap of string segments due to multiple parton interactions in pp or multiple nucleon interactions dominating in  nucleus\mbox{--}nucleus collisions.
EPOS-LHC reproduces the increased baryon-to-meson ratios at intermediate \pT~as a consequence of radial flow in high-multiplicity pp events~\cite{ALICE_7TeV_multiplicity}. 
Both PYTHIA8 and EPOS-LHC are tuned to reproduce the charged particle multiplicity and the production of identified hadrons (such as \pion, \K, p, \rmLambda, \rmXi) measured in pp collisions at \sqrtS~=~7~TeV~\cite{EPOS}.

Figure~\ref{Model_comparison_total} shows the comparison of the measured \simplekstarch~\pT~spectra at \sqrtS~=~5.02, 8, and 13~TeV with the PYTHIA6 (Perugia 2011 tune)~\cite{Perugia2011} and the PYTHIA8 (Monash 2013 tune) generators~\cite{Monash2013}, and EPOS-LHC~\cite{EPOS}. The bottom panels show the ratios of the model predictions to the measured distributions for \simplekstarch~mesons. 
The agreement with data improves with the collision energy. The best agreement is reached with PYTHIA6-Perugia 2011 and PYTHIA8-Monash 2013 for 13 TeV collisions.
None of the models considered for comparison is able to fully reproduce the data. For all three energies the models overestimate by a factor of 1.5\mbox{--}2 the yield for \pT~$<$~0.5~\gmom~and underestimate it in the intermediate \pT~region. EPOS-LHC predictions largely overestimate the data in the high-\pT~region, whereas an agreement within the uncertainties is observed for PYTHIA6 and also for PYTHIA8 at $\sqrt{s}~=~13$~TeV. Agreement is also observed with PYTHIA6 for \pT~$>$~4~\gmom~at $\sqrt{s}~=~8$~TeV. These results complement the observation reported in Ref.~\cite{ALICE_particle_13TeV} confirming that a more accurate tuning of the models is needed to reproduce the phase-space distribution of strange hadrons.

An evolution of the transverse momentum spectra with the collision energy is clearly observed in the left panel of Fig.~\ref{Energy_dependence}, where the ratios of the \simplekstarch~transverse-momentum spectra at 
\sqrtS~$=$~8 and 13~TeV to the one at \sqrtS~$=$~5.02~TeV are reported. 
The systematic uncertainties associated with the estimate of the material budget of the ALICE detector and the hadronic interaction cross section used in the simulations are the same for the different collision energies. Hence, they cancel out in the propagation of the uncertainties to the ratio.  
For \pT~$>$~1~\gmom, a hardening of the \simplekstarch~\pT~spectrum is observed from 5.02 to 13~TeV, which is indicative of an increasing contribution of hard scattering processes in particle production with the collision energy. 
In the right panel of Fig.~\ref{Energy_dependence}~the ratios  of the \Kp~+~\Km and \pionp~+~\pionm~\pT~distributions at 
\sqrtS~$=$~13~TeV~\cite{ALICE_particle_13TeV} to the ones at \sqrtS~$=$~5.02~TeV~\cite{ALICE_particle_5TeV} are compared to the same ratio for \simplekstarch. Distributions of these ratios are similar for the different particle species as shown in ref.~\cite{ALICE_particle_13TeV} for ratios of \pT~distributions at \sqrtS~$=$~13~TeV to the one at \sqrtS~$=$~7~TeV. 
These distributions, like the ones for \simplekstarch, show a progressive and significant evolution of the spectral shape at high \pT~with increasing collision energy and the shape independent of \pT~within uncertainties in the soft regime, \pT~$<$~1~\gmom. 
 
In the left panel of Fig.~\ref{Energy_dependence} the ratios of the \simplekstarch~transverse-momentum spectra at \sqrtS~$=$~8 and 13~TeV to the one at \sqrtS~$=$~5.02~TeV predicted by PYTHIA6, PYTHIA8 and EPOS-LHC are also shown. PYTHIA6 and PYTHIA8 predict a larger hardening with the energy, while EPOS-LHC is consistent with data.

\subsection{Energy dependence of \Bdndy, \BmeanpT~and \Bsimplekstarch/\bf{K}~ratio}
\label{sec:Kcharg_K0}
The measurements of particle production and particle ratios in pp collisions are important, also as a baseline for comparison with heavy-ion reactions. 
The per-event \pT-integrated \simplekstarch~yields (corresponding to $1/{N_{\mathrm{INEL}}} \times $ \dndy, hereby denoted as \dndy~for brevity)~for inelastic collisions and the mean transverse momenta \meanpT~are determined by integrating and averaging the transverse momentum spectra over the measured range and are listed in Tab.~\ref{dndy}.
For per-event~\pT-integrated yields and \meanpT~statistical uncertainties are estimated varying the data randomly inside the estimated uncertainties of each bin. The systematic uncertainties are computed assuming a full correlation across \pT. The uncertainty on \dndy~is estimated from the highest and lowest spectra allowed by the bin-by-bin systematic uncertainties whereas in  the case of the \meanpT~the allowed hardest and the softer \pT~distribution are considered. 

\begin{table*}
\caption{The per-event~\pT-integrated ({\mbox{\K$^{*+}$}} + {\mbox{\K$^{*-}$}})/2 yield for inelastic events in the interval 0~$<$~\pT~$<$~15~\gmom~at midrapidity, \dndy, the mean transverse momentum, \meanpT, and \simplekstarch/\K~for inelastic pp collisions at \sqrtS~=~5.02, 8 and 13~TeV. The kaon yield is (\K$^+$~+~\K$^-$)/2~\cite{ALICE_particle_5TeV,ALICE_Kstar_8TeV,ALICE_particle_13TeV}. The first uncertainty is statistical and the second one is the systematic uncertainty. The systematic uncertainty on \dndy~due to the normalization to inelastic collisions (2.51\%, 2.72\% and 2.55\% for 5.02, 8, and 13~TeV, respectively) is not included.}
\begin{center}
\begin{tabular}{cccc}
\hline\noalign{\smallskip}
\sqrtS~(TeV) & \dndy & \meanpT~(\gmom) & \simplekstarch/\K\\
\noalign{\smallskip}\hline
\hline
5.02 & 0.095 $\pm$ 0.001 $\pm$ 0.006 & 1.04 $\pm$ 0.01 $\pm$ 0.02 & 0.35 $\pm$ 0.01 $\pm$ 0.02\\
8 & 0.106 $\pm$ 0.002 $\pm$ 0.008 & 1.08 $\pm$ 0.02 $\pm$ 0.02 & 0.34 $\pm$ 0.01 $\pm$ 0.03\\
13 & 0.108~$\pm$~0.002~$\pm$~0.007 & 1.15~$\pm$~0.02~$\pm$~0.02 & 0.35~$\pm$~0.01~$\pm$~0.03\\
\noalign{\smallskip}\hline\hline
\end{tabular}
\end{center}
\label{dndy}       
\end{table*}

The per-event~\pT-integrated yield of the \simplekstarch~in inelastic pp collisions increases from \sqrtS~=~5.02~TeV to 13~TeV by 13.5~$\pm$~1.2$\%$. 
The hardening of the \simplekstarch~transverse momentum spectra reported in 
Fig.~\ref{Energy_dependence} manifests itself in the increasing mean transverse momentum.  
In pp collisions, the measured \meanpT~at $\sqrt{s}~=$~13~TeV is 11.1 $\pm$ 0.3$\%$ larger than at $\sqrt{s}~=~5.02$~TeV. Similar increasing trend of per-event~\pT-integrated yields and mean \pT~are observed for \simplekstarZ~across the same collisions 
energies~\cite{ALICE_Kstar_5TeV,ALICE_Kstar_8TeV,ALICE_particle_13TeV}.

Using the \simplekstarch~yields presented in this paper and the long-lived \K$^{\pm}$~production measured by ALICE at the same pp collision energies~\cite{ALICE_particle_5TeV,ALICE_Kstar_8TeV,ALICE_particle_13TeV}, the values of the 
\simplekstarch/\K~ratio were estimated and reported in Tab.~\ref{dndy}. 
The value of \dndy~for (\K$^+$+\K$^-$)~in \pp~collisions at \sqrtS~$=$~8~TeV was estimated by fitting the data points at \sqrtS~=~0.9, 2.76 and 7~TeV~\cite{ALICE_Kstar_8TeV} with the polynomial function $A(\sqrt s)^n +B$, where $A$, $n$ and $B$ are the fit parameters and by extrapolating the value for \sqrtS~=~8~TeV. 
Due to the fact that the same data samples were analyzed to extract both resonance and kaon yields, the uncertainties due to the absolute normalization cancel and therefore they are not included in the systematic uncertainties of these ratios. Consistent values are obtained for the ratio at the three collision energies.  
These ratios are presented in Fig.~\ref{Kstar_K_vs_energy}~together with the results obtained for \simplekstarZ/\K~in different collisions at 
RHIC~\cite{STAR_Kstar_200,STAR_Kstar_CuCu_AuAu,STAR_Kstar_dAu} and LHC~\cite{Kstar_PbPb,Kstar_7TeV,Kstar_pPb,Kstar_highpt,ALICE_particle_7TeV,ALICE_particle_276Pb,ALICE_Kstar_5TeV,
ALICE_Kstar_8TeV,ALICE_particle_13TeV,ALICE_particle_5TeV} energies. The \simplekstarch/K ratios predicted by PYTHIA6-Perugia 2011~\cite{Perugia2011}, PYTHIA8-Monash 2013~\cite{Monash2013} and EPOS-LHC~\cite{EPOS} at 5.02, 8 and 13~TeV are reported in Fig.~\ref{Kstar_K_vs_energy} with dashed lines. The predicted ratios do not change varying the collision energy and are in agreement with the measured values within uncertainties.     
In pp, \mbox{p--A}~and \mbox{d--A}~collisions at RHIC and the LHC, the \rsimplekstar/\K~ratio do not exhibit a strong dependence on the colliding system size or the centre-of-mass energy.
A lower value is reported for \simplekstarZ/\K~ratio in central \mbox{A--A} collisions both at RHIC and LHC energies. The observed suppression of the \simplekstarZ/\K~ratio is currently understood as the result of re-scattering and regeneration effects in the hadronic phase of heavy-ion collisions, with the former dominating over the latter~\cite{Kstar_PbPb,Kstar_highpt,ALICE_Kstar_5TeV}.

\begin{figure*}[h]
\centering
\includegraphics[width= 0.70\textwidth]{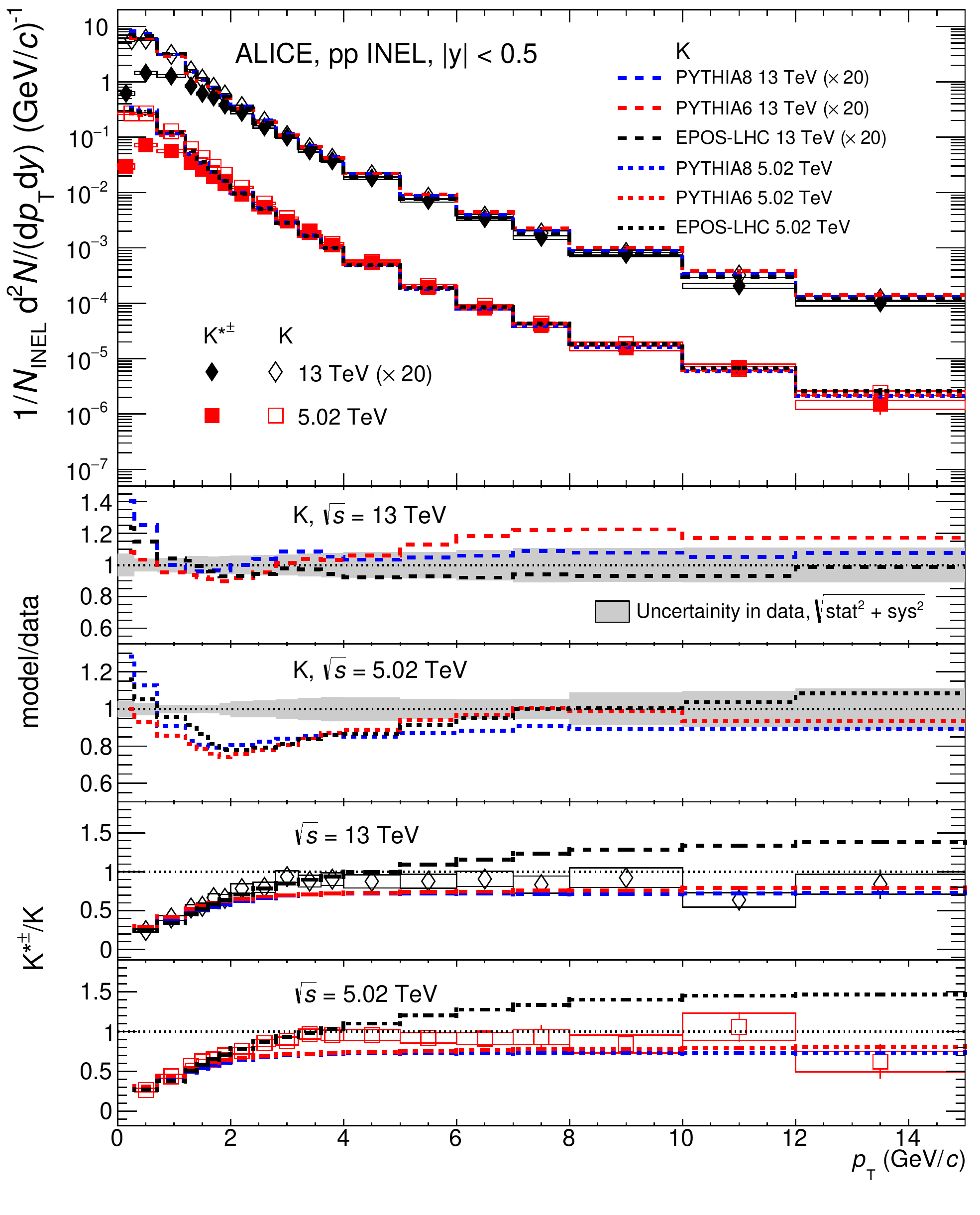}
\caption{(Colour online) (Upper panel) The \pT~spectra of \simplekstarch~in inelastic pp collisions at \sqrtS~=~5.02 and 13~TeV (full symbols) are compared to the \pT~spectra~of \Kpm~mesons (open symbols) at the same energies~\cite{ALICE_particle_5TeV,ALICE_particle_13TeV}. Statistical and systematic uncertainties are reported as error bars and boxes, respectively. Red, blue and black lines represent the \K~spectra predicted with PYTHIA6-Perugia 2011~\cite{Perugia2011}, PYTHIA8-Monash 2013~\cite{Monash2013} and EPOS-LHC~\cite{EPOS}, respectively. (Middle panels) The ratios of the rebinned predictions to the measured \pT~distributions for \Kpm~are reported in the two middle panels. The shaded bands represent the fractional uncertainties of the data points. (Bottom panels) The ratio of each measured \pT~distribution for \simplekstarch~mesons at \sqrtS~=~5.02 (red points) and 13~TeV (black points) to the \K~spectrum at the same collision energy is reported in the bottom panels. Red, blue and black lines represent the \simplekstarch/K ratio predicted with PYTHIA6-Perugia 2011~\cite{Perugia2011}, PYTHIA8-Monash 2013~\cite{Monash2013} and EPOS-LHC~\cite{EPOS}, respectively. 
} 
\label{KStarPM_Kaon_pT} 
\end{figure*}

In the upper panel of Fig.~\ref{KStarPM_Kaon_pT} the \simplekstarch~and \Kpm~\cite{ALICE_particle_5TeV,ALICE_particle_13TeV} \pT~spectra at \sqrtS~=~5.02 and 13~TeV are compared.
At both energies the \Kpm~and the \simplekstarch~ spectra exhibit the same slopes and consistent yields for \pT~$>$~3~\gmom. For \pT~$<$~2~\gmom~a larger yield for \Kpm~is measured with respect to \simplekstarch.
In the same figure the \Kpm~\pT~spectra are compared with the PYTHIA6 (Perugia 2011 tune)~\cite{Perugia2011}, PYTHIA8 (Monash 2013 tune)~\cite{Monash2013} and EPOS-LHC~\cite{EPOS} generators. The ratios of the rebinned predictions to the measured \pT~distributions for \Kpm~are reported in the two middle panels. Likewise \simplekstarch, for \Kpm~the agreement with data improves at higher collision energies. The best agreement is reached for 13~TeV collisions. For both energies PYTHIA8 and EPOS-LHC overestimate by a factor of 1.3\mbox{--}1.4 the \K~yield for \pT~$<$~0.5~\gmom~while PYTHIA6-Perugia 2011 reproduces or slightly underestimates the spectra in the same region. At 5.02~TeV all the models underestimate the spectra in the 
1~$<$~\pT~$<$~~6~\gmom~region. For \pT~larger than 5~\gmom~PYTHIA6-Perugia2011 model at 13~TeV is not able to reproduce the \K~data by a factor 1.2.

The \pT~dependence of the \simplekstarch/\K~ratios for pp collisions at \sqrtS~=~5.02 and 13~TeV is shown in the bottom panels of Fig.~\ref{KStarPM_Kaon_pT}. These ratios increase at low \pT~and saturate for \pT~$>$~3.0~\gmom. The \simplekstarch/\K~ratios predicted by PYTHIA6, PYTHIA8 and EPOS-LHC are also shown for comparison.
While PYTHIA6 and PYTHIA8 slightly underestimate the ratios for \pT~larger than 2~\gmom, EPOS-LHC predictions largely overestimate the data in the high-\pT~region. All the generators describe rather well the distributions at low transverse momentum.  

\section{Summary}
\label{sec:conclusions}

The first measurements of the \simplekstarch~resonance in inelastic pp collisions at different (5.02, 8, and 13~TeV) LHC energies were presented. 
The transverse momentum spectra were measured at midrapidity in the range 0~$<$~\pT~$<$~15~\gmom~and \pT-integrated yields as well as \meanpT~were calculated. These measurements complement and confirm the previous results for \simplekstarZ~although with smaller systematic uncertainties. 

The ratios of the \simplekstarch~\pT~distributions at \sqrtS~=~8~TeV and 13~TeV to those at 5.02~TeV reveal a hardening of the spectra with increasing collision energy for \pT~$>$~1~\gmom.
An increase in \meanpT~by about 11\%~is observed going from \sqrtS~=~5.02 to 13~TeV. This is consistent with the expectation that the contribution of hard processes to particle production increases with the collision energy. 
The weak energy dependence of the spectra below 1~\gmom~is consistent with the relatively small increase of the yields, since the \pT-integrated yields are dominated by the low-\pT~part of the spectrum. A similar evolution of the ratios of the \pT~distributions at \sqrtS~$=$~13~TeV to the one at \sqrtS~$=$~5.02~TeV is observed for \Kp~+~\Km~and \pionp~+~\pionm. This confirms the independence of the evolution of the spectral shape from particle species as observed in~\cite{ALICE_particle_13TeV}.

At \sqrtS~=~5.02 and 13~TeV the \Kpm~and the \simplekstarch~spectra exhibit the same slopes and consistent yields for \pT~$>$~3~\gmom. This indicates that production mechanisms as gluon fragmentation should have the same importance in the generation of ground and excitated status of \K. 
Moreover the \simplekstarch/\K~\pT-integrated yield ratios for the three reported energies are equal within uncertainties. This confirms, with a smaller uncertainty, the independence of \rsimplekstar/\K~ratio in pp collisions at LHC energies and the weak dependence on the colliding system size or the centre-of-mass energy in pp, \mbox{p--A}~and \mbox{d--A}~collisions at RHIC and the LHC.

Predictions of QCD-inspired (PYTHIA6, PYTHIA8) and hybrid (EPOS-LHC) event generators are not able to fully describe the \simplekstarch~transverse momentum spectra. The ability of the models to both qualitatively and quantitatively describe the data improves with the collision energy. The best agreement is obtained with PYTHIA6-Perugia 2011 and PYTHIA8-Monash 2013 for 13~TeV. However, EPOS-LHC better reproduces the relative hardening of the \pT~spectrum with increasing collision energy. The \simplekstarch/\K~ratios predicted from the event generators are in agreement with the measured ones and, like in  data, are independent from the collision energy. All the generators describe reasonably well the \simplekstarch/\K~ratio measured at low \pT~while they fail for \pT~larger than 2~\gmom.


\newenvironment{acknowledgement}{\relax}{\relax}
\begin{acknowledgement}
\section*{Acknowledgements}

The ALICE Collaboration would like to thank all its engineers and technicians for their invaluable contributions to the construction of the experiment and the CERN accelerator teams for the outstanding performance of the LHC complex.
The ALICE Collaboration gratefully acknowledges the resources and support provided by all Grid centres and the Worldwide LHC Computing Grid (WLCG) collaboration.
The ALICE Collaboration acknowledges the following funding agencies for their support in building and running the ALICE detector:
A. I. Alikhanyan National Science Laboratory (Yerevan Physics Institute) Foundation (ANSL), State Committee of Science and World Federation of Scientists (WFS), Armenia;
Austrian Academy of Sciences, Austrian Science Fund (FWF): [M 2467-N36] and Nationalstiftung f\"{u}r Forschung, Technologie und Entwicklung, Austria;
Ministry of Communications and High Technologies, National Nuclear Research Center, Azerbaijan;
Conselho Nacional de Desenvolvimento Cient\'{\i}fico e Tecnol\'{o}gico (CNPq), Financiadora de Estudos e Projetos (Finep), Funda\c{c}\~{a}o de Amparo \`{a} Pesquisa do Estado de S\~{a}o Paulo (FAPESP) and Universidade Federal do Rio Grande do Sul (UFRGS), Brazil;
Ministry of Education of China (MOEC) , Ministry of Science \& Technology of China (MSTC) and National Natural Science Foundation of China (NSFC), China;
Ministry of Science and Education and Croatian Science Foundation, Croatia;
Centro de Aplicaciones Tecnol\'{o}gicas y Desarrollo Nuclear (CEADEN), Cubaenerg\'{\i}a, Cuba;
Ministry of Education, Youth and Sports of the Czech Republic, Czech Republic;
The Danish Council for Independent Research | Natural Sciences, the VILLUM FONDEN and Danish National Research Foundation (DNRF), Denmark;
Helsinki Institute of Physics (HIP), Finland;
Commissariat \`{a} l'Energie Atomique (CEA) and Institut National de Physique Nucl\'{e}aire et de Physique des Particules (IN2P3) and Centre National de la Recherche Scientifique (CNRS), France;
Bundesministerium f\"{u}r Bildung und Forschung (BMBF) and GSI Helmholtzzentrum f\"{u}r Schwerionenforschung GmbH, Germany;
General Secretariat for Research and Technology, Ministry of Education, Research and Religions, Greece;
National Research, Development and Innovation Office, Hungary;
Department of Atomic Energy Government of India (DAE), Department of Science and Technology, Government of India (DST), University Grants Commission, Government of India (UGC) and Council of Scientific and Industrial Research (CSIR), India;
Indonesian Institute of Science, Indonesia;
Istituto Nazionale di Fisica Nucleare (INFN), Italy;
Institute for Innovative Science and Technology , Nagasaki Institute of Applied Science (IIST), Japanese Ministry of Education, Culture, Sports, Science and Technology (MEXT) and Japan Society for the Promotion of Science (JSPS) KAKENHI, Japan;
Consejo Nacional de Ciencia (CONACYT) y Tecnolog\'{i}a, through Fondo de Cooperaci\'{o}n Internacional en Ciencia y Tecnolog\'{i}a (FONCICYT) and Direcci\'{o}n General de Asuntos del Personal Academico (DGAPA), Mexico;
Nederlandse Organisatie voor Wetenschappelijk Onderzoek (NWO), Netherlands;
The Research Council of Norway, Norway;
Commission on Science and Technology for Sustainable Development in the South (COMSATS), Pakistan;
Pontificia Universidad Cat\'{o}lica del Per\'{u}, Peru;
Ministry of Education and Science, National Science Centre and WUT ID-UB, Poland;
Korea Institute of Science and Technology Information and National Research Foundation of Korea (NRF), Republic of Korea;
Ministry of Education and Scientific Research, Institute of Atomic Physics and Ministry of Research and Innovation and Institute of Atomic Physics, Romania;
Joint Institute for Nuclear Research (JINR), Ministry of Education and Science of the Russian Federation, National Research Centre Kurchatov Institute, Russian Science Foundation and Russian Foundation for Basic Research, Russia;
Ministry of Education, Science, Research and Sport of the Slovak Republic, Slovakia;
National Research Foundation of South Africa, South Africa;
Swedish Research Council (VR) and Knut \& Alice Wallenberg Foundation (KAW), Sweden;
European Organization for Nuclear Research, Switzerland;
Suranaree University of Technology (SUT), National Science and Technology Development Agency (NSDTA) and Office of the Higher Education Commission under NRU project of Thailand, Thailand;
Turkish Energy, Nuclear and Mineral Research Agency (TENMAK), Turkey;
National Academy of  Sciences of Ukraine, Ukraine;
Science and Technology Facilities Council (STFC), United Kingdom;
National Science Foundation of the United States of America (NSF) and United States Department of Energy, Office of Nuclear Physics (DOE NP), United States of America.
\end{acknowledgement}

\bibliographystyle{utphys}   
\bibliography{mybib}

\newpage
\appendix

%
%

\section{The ALICE Collaboration}
\label{app:collab}
%
\begingroup
\small
\begin{flushleft}
S.~Acharya$^{\rm 143}$, 
D.~Adamov\'{a}$^{\rm 98}$, 
A.~Adler$^{\rm 76}$, 
J.~Adolfsson$^{\rm 83}$, 
G.~Aglieri Rinella$^{\rm 35}$, 
M.~Agnello$^{\rm 31}$, 
N.~Agrawal$^{\rm 55}$, 
Z.~Ahammed$^{\rm 143}$, 
S.~Ahmad$^{\rm 16}$, 
S.U.~Ahn$^{\rm 78}$
% ...

\section*{Affiliation notes}

$^{\rm I}$ Deceased\\
$^{\rm II}$ Also at: Italian National Agency for New Technologies, Energy and Sustainable Economic Development (ENEA), Bologna, Italy\\
$^{\rm III}$ Also at: Dipartimento DET del Politecnico di Torino, Turin, Italy\\
$^{\rm IV}$ Also at: M.V. Lomonosov Moscow State University, D.V. Skobeltsyn Institute of Nuclear, Physics, Moscow, Russia\\
$^{\rm V}$ Also at: Institute of Theoretical Physics, University of Wroclaw, Poland\\

\section*{Collaboration Institutes}

$^{1}$ A.I. Alikhanyan National Science Laboratory (Yerevan Physics Institute) Foundation, Yerevan, Armenia\\
$^{2}$ AGH University of Science and Technology, Cracow, Poland\\
$^{3}$ Bogolyubov Institute for Theoretical Physics, National Academy of Sciences of Ukraine, Kiev, Ukraine\\
$^{4}$ Bose Institute, Department of Physics  and Centre for Astroparticle Physics and Space Science (CAPSS), Kolkata, India\\
$^{5}$ Budker Institute for Nuclear Physics, Novosibirsk, Russia\\
$^{6}$ California Polytechnic State University, San Luis Obispo, California, United States\\
$^{7}$ Central China Normal University, Wuhan, China\\
$^{8}$ Centro de Aplicaciones Tecnol\'{o}gicas y Desarrollo Nuclear (CEADEN), Havana, Cuba\\
$^{9}$ Centro de Investigaci\'{o}n y de Estudios Avanzados (CINVESTAV), Mexico City and M\'{e}rida, Mexico\\
$^{10}$ Chicago State University, Chicago, Illinois, United States\\
$^{11}$ China Institute of Atomic Energy, Beijing, China\\
$^{12}$ Chungbuk National University, Cheongju, Republic of Korea\\
$^{13}$ Comenius University Bratislava, Faculty of Mathematics, Physics and Informatics, Bratislava, Slovakia\\
$^{14}$ COMSATS University Islamabad, Islamabad, Pakistan\\
$^{15}$ Creighton University, Omaha, Nebraska, United States\\
$^{16}$ Department of Physics, Aligarh Muslim University, Aligarh, India\\
$^{17}$ Department of Physics, Pusan National University, Pusan, Republic of Korea\\
$^{18}$ Department of Physics, Sejong University, Seoul, Republic of Korea\\
$^{19}$ Department of Physics, University of California, Berkeley, California, United States\\
$^{20}$ Department of Physics, University of Oslo, Oslo, Norway\\
$^{21}$ Department of Physics and Technology, University of Bergen, Bergen, Norway\\
$^{22}$ Dipartimento di Fisica dell'Universit\`{a} 'La Sapienza' and Sezione INFN, Rome, Italy\\
$^{23}$ Dipartimento di Fisica dell'Universit\`{a} and Sezione INFN, Cagliari, Italy\\
$^{24}$ Dipartimento di Fisica dell'Universit\`{a} and Sezione INFN, Trieste, Italy\\
$^{25}$ Dipartimento di Fisica dell'Universit\`{a} and Sezione INFN, Turin, Italy\\
$^{26}$ Dipartimento di Fisica e Astronomia dell'Universit\`{a} and Sezione INFN, Bologna, Italy\\
$^{27}$ Dipartimento di Fisica e Astronomia dell'Universit\`{a} and Sezione INFN, Catania, Italy\\
$^{28}$ Dipartimento di Fisica e Astronomia dell'Universit\`{a} and Sezione INFN, Padova, Italy\\
$^{29}$ Dipartimento di Fisica e Nucleare e Teorica, Universit\`{a} di Pavia, Pavia, Italy\\
$^{30}$ Dipartimento di Fisica `E.R.~Caianiello' dell'Universit\`{a} and Gruppo Collegato INFN, Salerno, Italy\\
$^{31}$ Dipartimento DISAT del Politecnico and Sezione INFN, Turin, Italy\\
$^{32}$ Dipartimento di Scienze e Innovazione Tecnologica dell'Universit\`{a} del Piemonte Orientale and INFN Sezione di Torino, Alessandria, Italy\\
$^{33}$ Dipartimento di Scienze MIFT, Universit\`{a} di Messina, Messina, Italy\\
$^{34}$ Dipartimento Interateneo di Fisica `M.~Merlin' and Sezione INFN, Bari, Italy\\
$^{35}$ European Organization for Nuclear Research (CERN), Geneva, Switzerland\\
$^{36}$ Faculty of Electrical Engineering, Mechanical Engineering and Naval Architecture, University of Split, Split, Croatia\\
$^{37}$ Faculty of Engineering and Science, Western Norway University of Applied Sciences, Bergen, Norway\\
$^{38}$ Faculty of Nuclear Sciences and Physical Engineering, Czech Technical University in Prague, Prague, Czech Republic\\
$^{39}$ Faculty of Science, P.J.~\v{S}af\'{a}rik University, Ko\v{s}ice, Slovakia\\
$^{40}$ Frankfurt Institute for Advanced Studies, Johann Wolfgang Goethe-Universit\"{a}t Frankfurt, Frankfurt, Germany\\
$^{41}$ Fudan University, Shanghai, China\\
$^{42}$ Gangneung-Wonju National University, Gangneung, Republic of Korea\\
$^{43}$ Gauhati University, Department of Physics, Guwahati, India\\
$^{44}$ Helmholtz-Institut f\"{u}r Strahlen- und Kernphysik, Rheinische Friedrich-Wilhelms-Universit\"{a}t Bonn, Bonn, Germany\\
$^{45}$ Helsinki Institute of Physics (HIP), Helsinki, Finland\\
$^{46}$ High Energy Physics Group,  Universidad Aut\'{o}noma de Puebla, Puebla, Mexico\\
$^{47}$ Hiroshima University, Hiroshima, Japan\\
$^{48}$ Hochschule Worms, Zentrum  f\"{u}r Technologietransfer und Telekommunikation (ZTT), Worms, Germany\\
$^{49}$ Horia Hulubei National Institute of Physics and Nuclear Engineering, Bucharest, Romania\\
$^{50}$ Indian Institute of Technology Bombay (IIT), Mumbai, India\\
$^{51}$ Indian Institute of Technology Indore, Indore, India\\
$^{52}$ Indonesian Institute of Sciences, Jakarta, Indonesia\\
$^{53}$ INFN, Laboratori Nazionali di Frascati, Frascati, Italy\\
$^{54}$ INFN, Sezione di Bari, Bari, Italy\\
$^{55}$ INFN, Sezione di Bologna, Bologna, Italy\\
$^{56}$ INFN, Sezione di Cagliari, Cagliari, Italy\\
$^{57}$ INFN, Sezione di Catania, Catania, Italy\\
$^{58}$ INFN, Sezione di Padova, Padova, Italy\\
$^{59}$ INFN, Sezione di Pavia, Pavia, Italy\\
$^{60}$ INFN, Sezione di Roma, Rome, Italy\\
$^{61}$ INFN, Sezione di Torino, Turin, Italy\\
$^{62}$ INFN, Sezione di Trieste, Trieste, Italy\\
$^{63}$ Inha University, Incheon, Republic of Korea\\
$^{64}$ Institute for Gravitational and Subatomic Physics (GRASP), Utrecht University/Nikhef, Utrecht, Netherlands\\
$^{65}$ Institute for Nuclear Research, Academy of Sciences, Moscow, Russia\\
$^{66}$ Institute of Experimental Physics, Slovak Academy of Sciences, Ko\v{s}ice, Slovakia\\
$^{67}$ Institute of Physics, Homi Bhabha National Institute, Bhubaneswar, India\\
$^{68}$ Institute of Physics of the Czech Academy of Sciences, Prague, Czech Republic\\
$^{69}$ Institute of Space Science (ISS), Bucharest, Romania\\
$^{70}$ Institut f\"{u}r Kernphysik, Johann Wolfgang Goethe-Universit\"{a}t Frankfurt, Frankfurt, Germany\\
$^{71}$ Instituto de Ciencias Nucleares, Universidad Nacional Aut\'{o}noma de M\'{e}xico, Mexico City, Mexico\\
$^{72}$ Instituto de F\'{i}sica, Universidade Federal do Rio Grande do Sul (UFRGS), Porto Alegre, Brazil\\
$^{73}$ Instituto de F\'{\i}sica, Universidad Nacional Aut\'{o}noma de M\'{e}xico, Mexico City, Mexico\\
$^{74}$ iThemba LABS, National Research Foundation, Somerset West, South Africa\\
$^{75}$ Jeonbuk National University, Jeonju, Republic of Korea\\
$^{76}$ Johann-Wolfgang-Goethe Universit\"{a}t Frankfurt Institut f\"{u}r Informatik, Fachbereich Informatik und Mathematik, Frankfurt, Germany\\
$^{77}$ Joint Institute for Nuclear Research (JINR), Dubna, Russia\\
$^{78}$ Korea Institute of Science and Technology Information, Daejeon, Republic of Korea\\
$^{79}$ KTO Karatay University, Konya, Turkey\\
$^{80}$ Laboratoire de Physique des 2 Infinis, Ir\`{e}ne Joliot-Curie, Orsay, France\\
$^{81}$ Laboratoire de Physique Subatomique et de Cosmologie, Universit\'{e} Grenoble-Alpes, CNRS-IN2P3, Grenoble, France\\
$^{82}$ Lawrence Berkeley National Laboratory, Berkeley, California, United States\\
$^{83}$ Lund University Department of Physics, Division of Particle Physics, Lund, Sweden\\
$^{84}$ Moscow Institute for Physics and Technology, Moscow, Russia\\
$^{85}$ Nagasaki Institute of Applied Science, Nagasaki, Japan\\
$^{86}$ Nara Women{'}s University (NWU), Nara, Japan\\
$^{87}$ National and Kapodistrian University of Athens, School of Science, Department of Physics , Athens, Greece\\
$^{88}$ National Centre for Nuclear Research, Warsaw, Poland\\
$^{89}$ National Institute of Science Education and Research, Homi Bhabha National Institute, Jatni, India\\
$^{90}$ National Nuclear Research Center, Baku, Azerbaijan\\
$^{91}$ National Research Centre Kurchatov Institute, Moscow, Russia\\
$^{92}$ Niels Bohr Institute, University of Copenhagen, Copenhagen, Denmark\\
$^{93}$ Nikhef, National institute for subatomic physics, Amsterdam, Netherlands\\
$^{94}$ NRC Kurchatov Institute IHEP, Protvino, Russia\\
$^{95}$ NRC \guillemotleft Kurchatov\guillemotright  Institute - ITEP, Moscow, Russia\\
$^{96}$ NRNU Moscow Engineering Physics Institute, Moscow, Russia\\
$^{97}$ Nuclear Physics Group, STFC Daresbury Laboratory, Daresbury, United Kingdom\\
$^{98}$ Nuclear Physics Institute of the Czech Academy of Sciences, \v{R}e\v{z} u Prahy, Czech Republic\\
$^{99}$ Oak Ridge National Laboratory, Oak Ridge, Tennessee, United States\\
$^{100}$ Ohio State University, Columbus, Ohio, United States\\
$^{101}$ Petersburg Nuclear Physics Institute, Gatchina, Russia\\
$^{102}$ Physics department, Faculty of science, University of Zagreb, Zagreb, Croatia\\
$^{103}$ Physics Department, Panjab University, Chandigarh, India\\
$^{104}$ Physics Department, University of Jammu, Jammu, India\\
$^{105}$ Physics Department, University of Rajasthan, Jaipur, India\\
$^{106}$ Physikalisches Institut, Eberhard-Karls-Universit\"{a}t T\"{u}bingen, T\"{u}bingen, Germany\\
$^{107}$ Physikalisches Institut, Ruprecht-Karls-Universit\"{a}t Heidelberg, Heidelberg, Germany\\
$^{108}$ Physik Department, Technische Universit\"{a}t M\"{u}nchen, Munich, Germany\\
$^{109}$ Politecnico di Bari and Sezione INFN, Bari, Italy\\
$^{110}$ Research Division and ExtreMe Matter Institute EMMI, GSI Helmholtzzentrum f\"ur Schwerionenforschung GmbH, Darmstadt, Germany\\
$^{111}$ Russian Federal Nuclear Center (VNIIEF), Sarov, Russia\\
$^{112}$ Saha Institute of Nuclear Physics, Homi Bhabha National Institute, Kolkata, India\\
$^{113}$ School of Physics and Astronomy, University of Birmingham, Birmingham, United Kingdom\\
$^{114}$ Secci\'{o}n F\'{\i}sica, Departamento de Ciencias, Pontificia Universidad Cat\'{o}lica del Per\'{u}, Lima, Peru\\
$^{115}$ St. Petersburg State University, St. Petersburg, Russia\\
$^{116}$ Stefan Meyer Institut f\"{u}r Subatomare Physik (SMI), Vienna, Austria\\
$^{117}$ SUBATECH, IMT Atlantique, Universit\'{e} de Nantes, CNRS-IN2P3, Nantes, France\\
$^{118}$ Suranaree University of Technology, Nakhon Ratchasima, Thailand\\
$^{119}$ Technical University of Ko\v{s}ice, Ko\v{s}ice, Slovakia\\
$^{120}$ The Henryk Niewodniczanski Institute of Nuclear Physics, Polish Academy of Sciences, Cracow, Poland\\
$^{121}$ The University of Texas at Austin, Austin, Texas, United States\\
$^{122}$ Universidad Aut\'{o}noma de Sinaloa, Culiac\'{a}n, Mexico\\
$^{123}$ Universidade de S\~{a}o Paulo (USP), S\~{a}o Paulo, Brazil\\
$^{124}$ Universidade Estadual de Campinas (UNICAMP), Campinas, Brazil\\
$^{125}$ Universidade Federal do ABC, Santo Andre, Brazil\\
$^{126}$ University of Cape Town, Cape Town, South Africa\\
$^{127}$ University of Houston, Houston, Texas, United States\\
$^{128}$ University of Jyv\"{a}skyl\"{a}, Jyv\"{a}skyl\"{a}, Finland\\
$^{129}$ University of Kansas, Lawrence, Kansas, United States\\
$^{130}$ University of Liverpool, Liverpool, United Kingdom\\
$^{131}$ University of Science and Technology of China, Hefei, China\\
$^{132}$ University of South-Eastern Norway, Tonsberg, Norway\\
$^{133}$ University of Tennessee, Knoxville, Tennessee, United States\\
$^{134}$ University of the Witwatersrand, Johannesburg, South Africa\\
$^{135}$ University of Tokyo, Tokyo, Japan\\
$^{136}$ University of Tsukuba, Tsukuba, Japan\\
$^{137}$ Universit\'{e} Clermont Auvergne, CNRS/IN2P3, LPC, Clermont-Ferrand, France\\
$^{138}$ Universit\'{e} de Lyon, CNRS/IN2P3, Institut de Physique des 2 Infinis de Lyon , Lyon, France\\
$^{139}$ Universit\'{e} de Strasbourg, CNRS, IPHC UMR 7178, F-67000 Strasbourg, France, Strasbourg, France\\
$^{140}$ Universit\'{e} Paris-Saclay Centre d'Etudes de Saclay (CEA), IRFU, D\'{e}partment de Physique Nucl\'{e}aire (DPhN), Saclay, France\\
$^{141}$ Universit\`{a} degli Studi di Foggia, Foggia, Italy\\
$^{142}$ Universit\`{a} di Brescia, Brescia, Italy\\
$^{143}$ Variable Energy Cyclotron Centre, Homi Bhabha National Institute, Kolkata, India\\
$^{144}$ Warsaw University of Technology, Warsaw, Poland\\
$^{145}$ Wayne State University, Detroit, Michigan, United States\\
$^{146}$ Westf\"{a}lische Wilhelms-Universit\"{a}t M\"{u}nster, Institut f\"{u}r Kernphysik, M\"{u}nster, Germany\\
$^{147}$ Wigner Research Centre for Physics, Budapest, Hungary\\
$^{148}$ Yale University, New Haven, Connecticut, United States\\
$^{149}$ Yonsei University, Seoul, Republic of Korea\\

\bigskip 

\end{flushleft} 
\endgroup  
\end{document}